\documentclass[a4paper,11pt]{article}
\pdfoutput=1 

\usepackage{jcappub}
\usepackage{aas_macros}
\usepackage[cp1252]{inputenc}
\usepackage[T1]{fontenc} 
\usepackage{enumitem}

\newcommand\numberthis{\addtocounter{equation}{1}\tag{\theequation}}
\renewcommand{\vec}[1]{\mathbf{#1}}

\title{\boldmath A fully Lagrangian, non-parametric bias model for dark matter halos}

\author[a,1]{Xiaohan Wu,\note{Corresponding author.}}
\author[a]{Julian B.~Mu\~noz,}
\author[a]{Daniel Eisenstein}

\affiliation[a]{Harvard-Smithsonian Center for Astrophysics, 60 Garden Street, Cambridge 02138, MA, USA}

\emailAdd{xiaohan.wu@cfa.harvard.edu}
\emailAdd{julianmunoz@cfa.harvard.edu}
\emailAdd{deisenstein@cfa.harvard.edu}

\abstract{We present a non-parametric Lagrangian biasing model and fit the ratio of the halo and mass densities at the field level using the mass-weighted halo field in the \textsc{AbacusSummit} simulations at $z=0.5$.  Unlike the perturbative halo bias model that has been widely used in interpreting the observed large-scale structure traced by galaxies, we find a non-negative halo-to-mass ratio that increases monotonically with the linear overdensity $\delta_1$ in the initial Lagrangian space.  The bias expansion, however, does not guarantee non-negativity of the halo counts, and may lead to rising halo number counts at negative overdensities.  The shape of the halo-to-mass ratio is unlikely to be described by a polynomial function of $\delta_1$ and other quantities.  Especially for massive halos with $6\times10^{12}\ h^{-1}\ M_\odot$, the halo-to-mass ratio starts soaring up at $\delta_1>0$, substantially different from the predictions of the bias expansion.  We show that for the halo masses we consider ($M>3\times10^{11}\ h^{-1}\ M_\odot$) a non-parametric halo-to-mass ratio as a function of $\delta_1$ and its local derivative $\nabla^2\delta_1$ can recover the halo power spectra to sub-percent level accuracy for wavenumbers $k=0.01-0.1\ h\ {\rm Mpc}^{-1}$ given a proper smoothing scale to filter the initial density field, even though we do not fit the power spectrum directly.  However, there is mild dependence of the recovery of the halo power spectrum on the smoothing scale and other input parameters.  At $k<0.01\ h\ {\rm Mpc}^{-1}$ and for massive halos with $M>6\times10^{12}\ h^{-1}\ M_\odot$, our non-parametric model leads to a few percent overestimation of the halo power spectrum, indicating the need for larger or multiple smoothing scales.  The halo-to-mass ratios obtained qualitatively agree with intuitions from extended Press-Schechter theory.  We compare our framework to the bias expansion and discuss possible extensions.}

\begin{document}
\maketitle
\flushbottom

\section{Introduction}
\label{sec:intro}

The large-scale structure (LSS) of the universe has become a powerful tool for research in cosmology, providing information complementary to or inaccessible by the cosmic-microwave background \cite{planck18}.  Over the next decade, LSS surveys such as DESI \cite{desi1}, {\it Euclid} \cite{euclid11, euclid18}, and the LSST \cite{lsst} will generate vast amounts of cosmological data, providing stringent tests of our understanding of the universe.
All these surveys target biased tracers (e.g.~galaxies) of the underlying matter field.
Modeling the connection between these tracers and the underlying matter density is thus key to extracting the maximal information about the universe from observational data.
In this work we develop a fully non-parametric framework to find the abundance of halos from the initial Lagrangian density field, going beyond the traditional perturbative approaches.

In the Lagrangian formalism, the Lagrangian-space (pre-advection-)halo overdensity $\delta_h$ at the initial time can be written as a function $f$ of the local Lagrangian matter overdensity $\delta$ and two terms that encode nonlocality:
$\nabla_i \nabla_j \Phi$ and $\nabla_i v_j$, where $\Phi$ is the gravitational potential and $\vec{v}$ is the peculiar velocity \citep[for a recent review, see e.g.][]{desjacques18}.
This function $f$ determines the weight that a fluid element carry, which is then advected to the final redshift to give the Eulerian halo density field.
Previous works \cite{matsubara08, mcdonald09, assassi14, vlah16} have suggested that the functional form
\begin{equation}
1+\delta_h = f(\delta, \nabla^2\delta, \mathcal{G}_2)
\label{eq:f_func}
\end{equation}
provides an accurate description of biased tracers, where $\mathcal{G}_2$ is the tidal operator (equation~\eqref{eq:G2}).
Traditionally, $f$ is Taylor expanded around $\delta=0$ and a series of bias coefficients $b_i$ are used to encode the response of small-scale halo formation physics to the large-scale structure \cite{fry93, matsubara08, vlah16}:
\begin{equation}
f \approx 1 + b_1 \delta + b_2 (\delta^2 - \langle \delta^2 \rangle) + b_{\mathcal{G}_2}(\mathcal{G}_2 - \langle \mathcal{G}_2 \rangle) + b_{\nabla^2} \nabla^2\delta + ...,
\label{eq:f_pt}
\end{equation}
where $\langle \cdot \rangle$ represents a spatial average.
The above formalism can also be written in Eulerian space, where the final-time Eulerian halo overdensity and matter overdensity are related via the bias expansion.
Since the standard Eulerian bias model has been shown to lead to larger errors at reproducing the observed halo field than the Lagrangian one \cite{roth11, schmittfull19, modi19}, we will focus on the Lagrangian picture of linking the halo field to the initial density field, but avoid a Taylor expansion of $f$.
We will advect the halo and initial density fields to lower redshifts non-perturbatively using N-body simulations, which is more computationally expensive than computing displacements using the Zel'dovich approximation \citep[as done in][]{schmittfull19, modi19} but more accurate \cite{modi20, kokron21, zennaro21, pelliban21}.

Previous studies have focused on the bias expansion approach of describing the large-scale halo field and evaluating the biases \cite{kaiser84, desjacques10, musso12, baldauf15, modi17, lazeyras16, lazeyras18, lazeyras19, lazeyras21}.
With various improvements developed over the years, the bias expansion has achieved broad success in describing summary statistics such as the galaxy power spectrum and bispectrum \cite{chan12, baldauf12, saito14, abidi18, fujita20, modi20, kokron21}.
The effectiveness of perturbation theory has also been evaluated at the field level \cite{schmittfull19, schmittfull20, modi19, barreira21}.
An important merit of the bias expansion is that one can get physical intuition of bias parameters from the peak-background split argument and obtain theoretical predictions for the biases \cite{bardeen86, mo96, sheth99}.
However, the bias expansion can yield an unphysical relation between the galaxy and matter fields.
Unphysicality can manifest via a non-positive-definite $f$, as well as enhanced biases for underdense regions.

\begin{figure}[tbp]
\centering
\includegraphics[width=\linewidth]{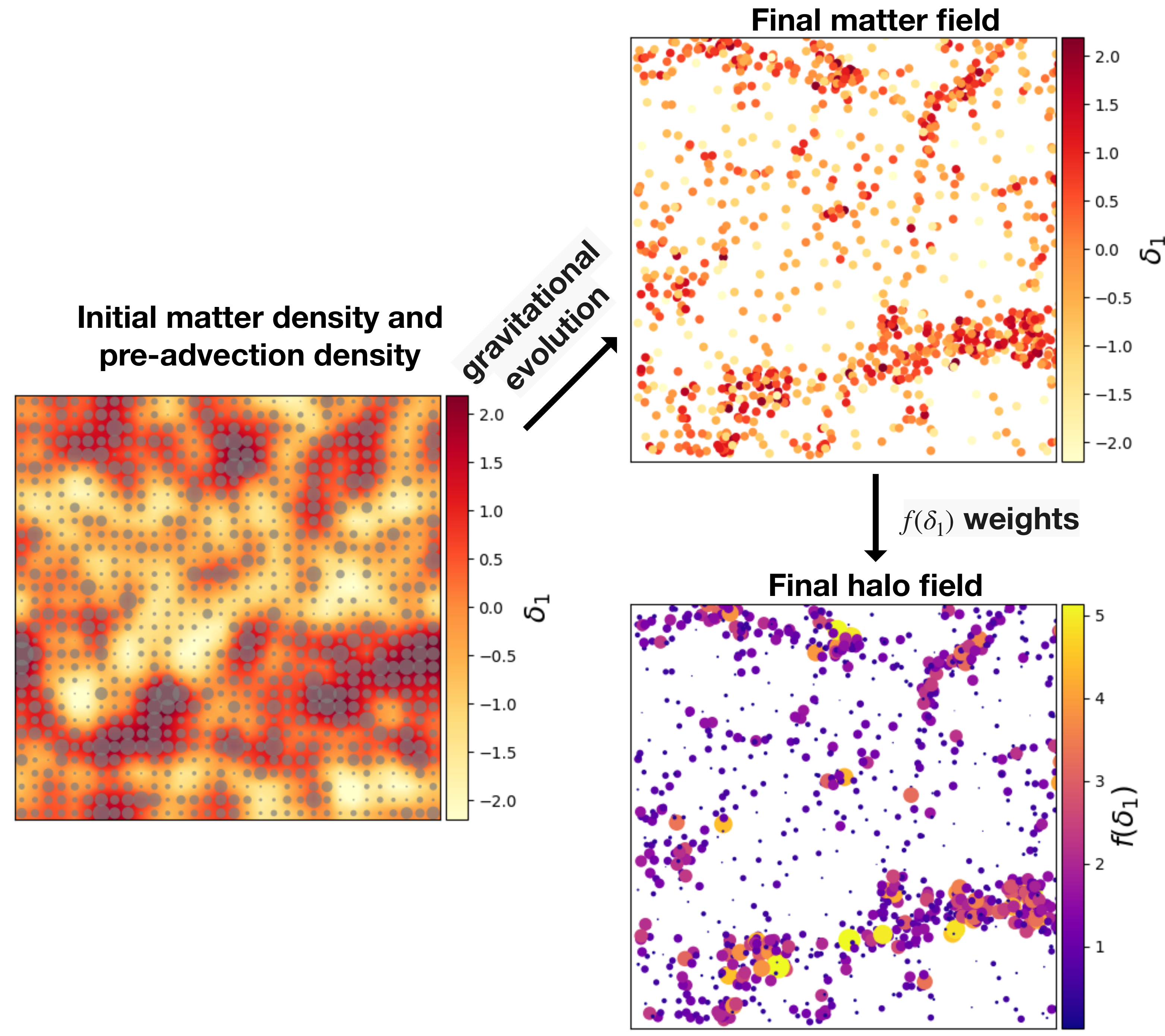}
\caption{\label{fig:illustration} A schematic illustration of our procedure to calculate a halo field given the $f$ weights that the particles in a simulation should carry.  Left panel shows the initial matter density field, illustrated by colors.  The overlying gray circles represent the $f$ weights that the particles carry, with the size of the circles showing the amplitude of $f$.  The top right panel illustrates the final matter field, where the particles shown by circles have been moved by gravity.  The bottom right panel presents the final halo field, where the circles follow the locations of the particles and the colors and sizes of the circles represent the amplitude of $f$.}
\end{figure}

Given these caveats of the bias expansion, we propose a fully Lagrangian, non-parametric halo bias model and measure the $f$ function in N-body simulations at the field level in real space.  We perform the measurements for mass-weighted halo fields, such that the resulting $f$ represents the halo-to-mass ratio that a patch in the initial Lagrangian space should carry to form halos at the final redshift.  Figure~\ref{fig:illustration} gives a schematic illustration of our procedure to calculate a halo field given the $f$ weights that the particles should carry.  We show that our non-parametric $f$ is non-negative by construction and monotonically increasing with density.  Its shape shows a clear deviation from a linear or quadratic function of the density, especially for more massive halos.  Our non-parametric $f$ leads to sub-percent level accuracy on the prediction of the halo power spectrum at $k\sim0.01-0.1\ h\ {\rm Mpc}^{-1}$ given an appropriate smoothing scale of the initial density field, albeit with a mild dependence on the smoothing scale and other input parameters.

The paper is organized as follows.  Section~\ref{sec:methods} introduces the formalism of our non-parametric $f$ and the simulations used.  Section~\ref{sec:results} shows $f$ measured for various halo mass cuts, the recovery of the halo power spectrum, and the dependencies on the parameters used.  We conclude in Section~\ref{sec:conclusions} and discuss possible extensions of our formalism.

\section{Methods}
\label{sec:methods}

\subsection{Lagrangian formalism of a non-parametric biasing model}

In the Lagrangian picture, a fluid element is mapped from its initial Lagrangian position $\vec{q}$ to its final Eulerian coordinates $\vec{x}$ at time $t$ through the displacement $\vec{\Psi}(\vec{q}, t)$:
\begin{equation}
\vec{x}(\vec{q}, t) = \vec{q} + \vec{\Psi}(\vec{q}, t).
\end{equation}
If the particles carry weights $f(\vec{q})$, then the resulting field at the final time $t$ can be obtained as $\int \mathrm{d}^3q \delta^D(\vec{x}-\vec{q}-\vec{\Psi}(\vec{q},t)) f(\vec{q})$, where $\delta^D$ denotes the 3-dimensional Dirac delta function.  Specifically, $f=1$ gives the Eulerian density field at time $t$.

The overall goal of this work is to find which weights $f$ lead to the correct halo field $1+\delta_h$ at time $t$.  We take $f$ to be a function of the smoothed linear overdensity $\delta_1$, its Laplacian $\nabla^2\delta_1$, and the corresponding tidal operator $\mathcal{G}_2$:
\begin{equation}
\mathcal{G}_2 = \sum_{ij} \left( \left[ \nabla_i \nabla_j \nabla^{-2} - \frac{1}{3}\delta^K_{ij} \right] \delta_1 \right)^2,
\label{eq:G2}
\end{equation}
where $\delta^K_{ij}$ is the Kronecker delta.
$\delta_1$ is defined as
\begin{equation}
\delta_1(\vec{q}) = \int \mathrm{d}^3q' W_{R_f}(|\vec{q}-\vec{q'}|) \delta(\vec{q'}),
\end{equation}
where $W_{R_f}$ is a Gaussian smoothing kernel of size $R_f$, and $\delta(\vec{q})$ is the unsmoothed linear overdensity.  We will explain the necessity of smoothing the density field later.  The final halo field is thus computed through
\begin{equation}
1+\delta_h = \int \mathrm{d}^3q \delta^D(\vec{x}-\vec{q}-\vec{\Psi}(\vec{q},t)) f(\delta_1, \nabla^2\delta_1, \mathcal{G}_2).
\label{eq:delta_model_displacement}
\end{equation}
Here the weights $f$ should satisfy
\begin{equation}
\int P(\delta_1, \nabla^2\delta_1, \mathcal{G}_2) f(\delta_1, \nabla^2\delta_1, \mathcal{G}_2) \mathrm{d}\delta_1 \mathrm{d}(\nabla^2\delta_1) \mathrm{d}\mathcal{G}_2 = 1,
\label{eq:inte_f_constraint}
\end{equation}
where $P(\delta_1, \nabla^2\delta_1)$ is the probability distribution of $(\delta_1, \nabla^2\delta_1, \mathcal{G}_2)$.

Instead of expanding $f$ in a Taylor series, we choose to fit a non-parametric $f$ using the initial conditions and the final halo field in N-body simulations.  To this end, we make a number $N_{\rm bins}$ of bins in the 3-dimensional volume of $\delta_1$-$\nabla^2\delta_1$-$\mathcal{G}_2$ and fit for the $f$ value within each bin.  We assign a weight $f$ to each dark matter particle at a final redshift $z$ according to the $(\delta_1, \nabla^2\delta_1, \mathcal{G}_2)$ values at its initial Lagrangian position.  We then grid the particles into $N_{\rm cells}$ grid cells by Cloud-In-Cell (CIC) interpolation using their locations at $z$ and the assigned weights $f$, which yields the predicted density field for the biased objects, $1 + \delta^{\rm model}_h$.  Comparing to the true halo density field $1 + \delta^{\rm true}_h$ obtained also with CIC interpolation and minimizing $\sum_j (\delta^{\rm model}_{h,j} - \delta^{\rm true}_{h,j})^2$ in real space gives the least squares solution to $f$, where $j$ denotes the grid index.
Figure~\ref{fig:illustration} gives a schematic illustration of how we calculate $\delta^{\rm model}_h$, where the particles carry their corresponding $f$ weights (gray circles in the left panel) given at the initial time to their final locations, forming the final halo field (bottom right panel).

We now briefly outline a mathematical derivation of $\delta^{\rm model}_h$.  At a location with grid index $j$, $\delta^{\rm model}_{h,j}$ is given by
\begin{equation}
1+\delta^{\rm model}_{h, j} = \sum_i w_{ij} f(\delta_{1,i}, \nabla^2\delta_{1,i}, \mathcal{G}_{2,i}),
\label{eq:delta_model_0}
\end{equation}
where $i$ denotes the indices of the dark matter particles, $w_{ij}$ is the CIC weight that the $i$-th particle contributes to the $j$-th grid point, and $f(\delta_{1,i}, \nabla^2\delta_{1,i}, \mathcal{G}_{2,i})$ is the weight that the $i$-th particle carries.  Suppose that $(\delta_{1,i}, \nabla^2\delta_{1,i}, \mathcal{G}_{2,i})$ falls in the $m$-th bin in the 3-dimensional $\delta_1$-$\nabla^2\delta_1$-$\mathcal{G}_2$ volume so that $f(\delta_{1,i}, \nabla^2\delta_{1,i}, \mathcal{G}_{2,i}) = f_m$, we then get
\begin{equation}
1+\delta^{\rm model}_{h, j} = \sum_m \sum_{i \in \mathcal{I}_m} w_{ij} f_m = \sum_m A_{jm} f_m,
\end{equation}
where $\mathcal{I}_m$ is the set of indices of particles that carry weight $f_m$.  $A$ is a $N_{\rm cells} \times N_{\rm bins}$ matrix whose $(j,m)$-th element is
\begin{equation}
A_{jm} = \sum_{i \in \mathcal{I}_m} w_{ij}.
\end{equation}
We thus aim to minimize the error of reproducing the true halo field by solving the quadratic optimization problem
\begin{align*}
&\operatorname*{argmin}_f\ \left( A f - \left( 1 + \delta^{\rm true}_h \right) \right)^T \left( Af - \left( 1 + \delta^{\rm true}_h \right) \right) \\
= &\operatorname*{argmin}_f\ f^T A^T A f - 2\left( 1 + \delta^{\rm true}_h \right)^T A f + {\rm const}.\numberthis
\label{eq:qp_objective}
\end{align*}

In practice, rather than using all information down to the pixel scale in the least-squares fit, we minimize $\sum_{k<k_{\rm max}} | \mathcal{F}(\delta^{\rm model}_h) - \mathcal{F}(\delta^{\rm true}_h) |^2$, where $\mathcal{F}(\cdot)$ denotes the Fourier transform and $k_{\rm max}$ is the maximum wavenumber that we sum up the residuals to.  Using Parseval's theorem, this sum of residuals in $k$-space can be written equivalently in real space as $\sum_j (\tilde{\delta}^{\rm model}_{h,j} - \tilde{\delta}^{\rm true}_{h,j})^2$, where the $\tilde{\delta}$'s are the real space halo fields filtered with a sharp-$k$ filter $W(k)$:
\begin{gather}
\tilde{\delta}^{\rm true}_h = \mathcal{F}^{-1}\left( \mathcal{F}\left( \delta^{\rm true}_h \right) W(k) \right) \\
\tilde{\delta}^{\rm model}_h = \mathcal{F}^{-1}\left( \mathcal{F}\left( \delta^{\rm model}_h \right) W(k) \right) = \sum_m \underbrace{\mathcal{F}^{-1}\left( \mathcal{F}\left( A_{*m} \right) W(k) \right)}_{\widetilde{A}_{*m}} f_m - 1.
\end{gather}
We thus still use equation~\eqref{eq:qp_objective} to calculate the objective function, but substituting $A$ and $\delta^{\rm true}_h$ with the filtered values $\widetilde{A}$ and $\tilde{\delta}^{\rm true}_h$.

Without constraints, the quadratic optimization 
problem of equation~\eqref{eq:qp_objective} can be easily solved with linear algebra.  However, we found that for some choices of the parameters $R_f, k_{\rm max},$ and $N_{\rm cells}$, the simple least-squares solution leads to negative $f$ in underdense regions, violating the physical intent of our formalism (see Section~\ref{sec:results_N150}).  The least-squares solution also does not guarantee the normalization constraint of equation~\eqref{eq:inte_f_constraint}.  We thus by default solve $f$ as a quadratic programming problem with the normalization constraint and the $f\ge0$ constraint using the Python package {\tt qpsolvers}\footnote{\url{https://github.com/stephane-caron/qpsolvers}}.  We will discuss how the results change with and without these constraints.

We note that although recent works such as \cite{schmittfull19, kokron21} do not smooth the initial density field as we do, the gridding of the field leads to an implicit smoothing on scales roughly corresponding to the cell size.  Our explicit smoothing leads to results that are independent of the grid size (but dependent on the smoothing scale).
Smoothing the initial field also makes the matrix $A^T A$ more diagonal.  Intuitively, smoothing with a large enough $R_f$ would lead to all particles at a point having the same weight $f$ and so each row of $A$ having only one non-zero element of 1, thus making $A^T A$ diagonal.
We choose to use a Gaussian smoothing kernel in this work.

In this work we use mass-weighted halos instead of number-weighted.  We are thus predicting the the ratio of the mass-weighted halo density to the total matter density.  We defer an examination against halos weighted by number or a halo occupation distribution model to future work.

\subsection{Simulations}

We now briefly outline the \textsc{AbacusSummit} simulations that we use in this work.
\textsc{AbacusSummit} \cite{maksimova21} is a suite of large, high-accuracy cosmological N-body simulations run with the Abacus N-body simulation code \cite{garrison18, garrison19, garrison21}.  Abacus utilizes a novel, fully disjoint split between the near-field and far-field gravitational sources, solving the former on GPU hardware and the latter with a variant of a multipole method \cite{metchnik09}.  The resulting code is both accurate and fast, up to 70M particle updates per second per node on Summit.

The \textsc{AbacusSummit} simulations were designed to meet and exceed the currently stated Cosmological Simulation Requirements of the Dark Energy Spectroscopic Instrument (DESI) survey \cite{desi1}.  We utilize a set of 25 simulations, each with $2\ h^{-1}$~Gpc box size and $6912^3$ particles, using the Planck2018 LCDM cosmology \cite{planck18}: $\Omega_{\rm m}=0.14237, h=0.6736, \sigma_8=0.807952$.  This gives a particle mass of $2\times10^9\ h^{-1}\ M_\odot$.  We use a force softening of $7.2h^{-1}$ proper kpc.

The initial conditions were generated at $z=99$ using the method proposed in \cite{garrison16}.  To obtain the $(\delta_1, \nabla^2\delta_1, \mathcal{G}_2)$ values associated with a particle, we interpolated the initial density field onto $1152^3$ grids and calculated the $(\delta_1, \nabla^2\delta_1, \mathcal{G}_2)$ values on each grid point given a smoothing scale $R_f$.  We then assign $(\delta_1, \nabla^2\delta_1, \mathcal{G}_2)$ values to each particle by looking for the nearest grid point to the particle's location in the initial space.

Halos are identified on the fly with the CompaSO Halo Finder which uses a hybrid FoF-SO algorithm (Hadzhiyska et al. submitted).  A kernel density estimate is first computed around all particles.  Particles with overdensity larger than 60 are then segmented with the FoF algorithm with linking length $0.25$ of the interparticle spacing.  Finally, halos are identified within each segmentation by a competitive spherical overdensity algorithm, with an overdensity threshold of 200.  Here we will only use halos at $z=0.5$ with at least 150 particles (corresponding to a halo mass of $M=3\times 10^{11}\,h^{-1}\,M_\odot$).

In addition to the $2\ h^{-1}$~Gpc simulations, on which we focus here, in Sec.~\ref{sec:results_smallbox} we use multiple $500\ h^{-1}$~Mpc small-box simulations with the same mass resolution and cosmological parameters as the large-box ones.

Given the increased computational cost incurred when fitting large-dimensional parameter spaces, we will study $f$ in the 2-dimensional planes $\delta_1$-$\nabla^2\delta_1$ and $\delta_1$-$\mathcal{G}_2$ separately, instead of fully exploring the 3-dimensional $\delta_1$-$\nabla^2\delta_1$-$\mathcal{G}_2$ volume.  We first make 40 bins in $\delta_1$ from $\delta_1/\sigma(\delta_1)=-4$ to 5, where $\sigma(\cdot)$ denotes the standard deviation.  In each bin of $\delta_1$, we make 5 bins in $\nabla^2\delta_1$ corresponding to $<5, 5-30, 30-70, 70-95, >95$ percentiles, or 5 bins in $\mathcal{G}_2$ representing $<10, 10-30, 30-70, 70-90, >90$ percentiles.

\begin{figure}[tbp]
\centering
\includegraphics[width=\linewidth]{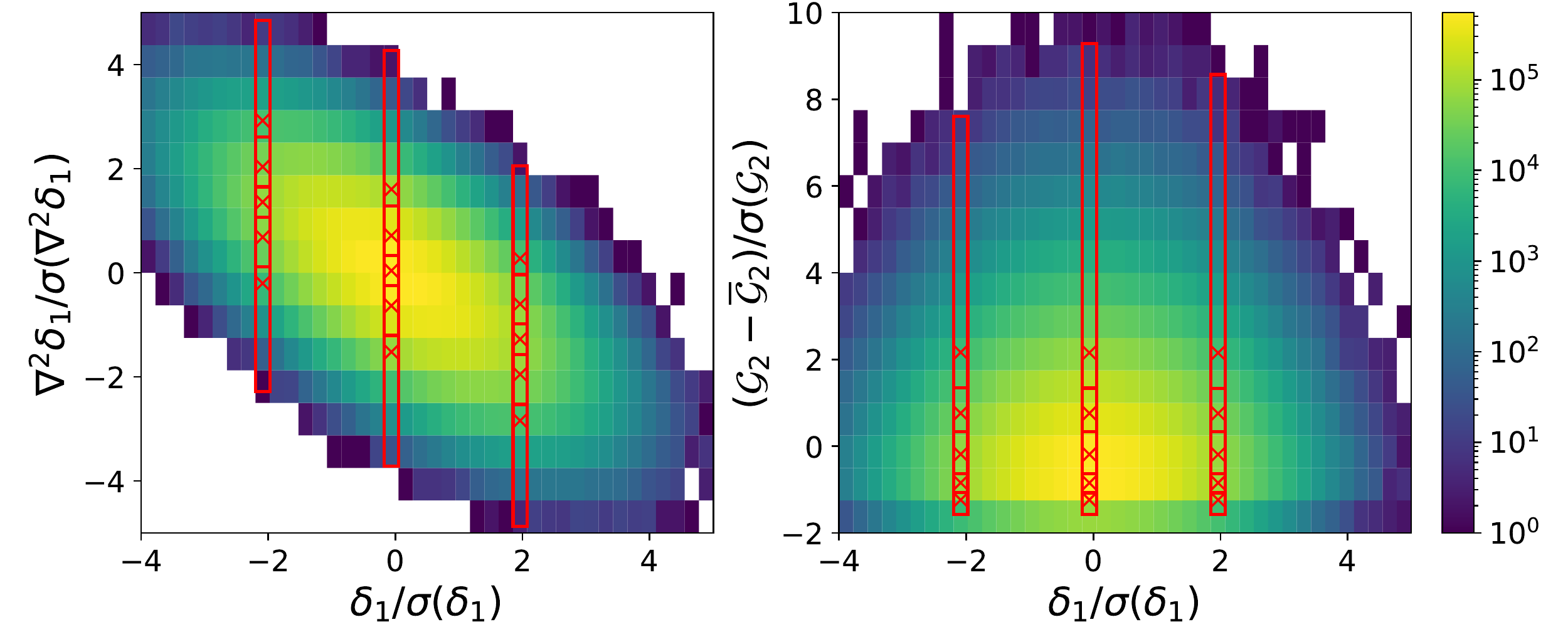}
\caption{\label{fig:2Dbins} An illustration of the bins used for fitting $f$.  Left and right panels show the 2D histograms of $\delta_1$ and $\nabla^2\delta_1$, and $\delta_1$ and $\mathcal{G}_2$ respectively, where the binning in $\delta_1$ correspond to the 40 bins we used for fitting $f$.  The vertical edges of the red rectangles show the boundaries of 3 different bins in $\delta_1$ centered around $\delta_1/\sigma(\delta_1)=-2,0,2$, and the horizontal edges represent the boundaries of bins in either $\nabla^2\delta_1$ or $\mathcal{G}_2$.  The red crosses illustrate the centers of the bins.}
\end{figure}

Figure~\ref{fig:2Dbins} illustrate the 2D histograms of $\delta_1$ and $\nabla^2\delta_1$ (left), and $\delta_1$ and $\mathcal{G}_2$ (right), where the bins in $\delta_1$ correspond to the 40 bins we used for fitting $f$.  The vertical edges of the red rectangles show the boundaries of 3 different bins in $\delta_1$ centered around $\delta_1/\sigma(\delta_1)=-2,0,2$, and the horizontal edges represent the boundaries of bins in either $\nabla^2\delta_1$ or $\mathcal{G}_2$.  The red crosses illustrate the centers of the bins.
We note that owing to the increasing computational cost, we are not able to make more bins in $\nabla^2\delta_1$ and $\mathcal{G}_2$, even though these bins are much sparser than those in $\delta_1$.  However, most bins appear compact, except the boundary bins corresponding to $\nabla^2\delta_1<5$ percentile or $>95$ percentile, and $\mathcal{G}_2>90$ percentile, where the particle numbers are small.  We have also verified that making 4 bins in $\nabla^2\delta_1$ instead of 5 bins does not impact our $f$ or recovery of the halo power spectrum, indicating that the binning is good enough for the purpose of this paper.

\section{The non-parametric halo bias model}
\label{sec:results}

We now discuss the non-parametric halo-to-mass ratio ($f$) solutions and how well they recover the halo power spectra for mass-weighted halos with $M>3\times10^{11} - 6\times10^{12}\ h^{-1}\ M_\odot$ at $z=0.5$.
Since we focus on the mass-weighted halo field, a proper value for the smoothing scale $R_f$ should be determined by enclosing the mass-weighted average mass of the halos.  The mass-weighted mean masses with $M>3\times10^{11}\ h^{-1}\ M_\odot$ and $6\times10^{12}\ h^{-1}\ M_\odot$ are $2.4\times10^{13}\ h^{-1}\ M_\odot$ and $5.1\times10^{13}\ h^{-1}\ M_\odot$ respectively, which correspond to Gaussian filters with $R_f=2.6\ h^{-1}$~Mpc and $3.3\ h^{-1}$~Mpc respectively.  We thus by default adopt a Gaussian smoothing scale $R_f=3\ h^{-1}$~Mpc and $k_{\rm max}=0.3\ h\ {\rm Mpc}^{-1} (\sim 1/R_f)$ to compute $f$ for all mass cuts.
We will discuss results with other $R_f$ and $k_{\rm max}$ choices later on.
We interpolate the particles to a $400^3$ grid and compute the matrix $A$ and the $f$ solution.  This gives a Nyquist frequency of $0.63\ h\ {\rm Mpc}^{-1}$ and a cell size of $5\ h^{-1}$~Mpc, larger than the biggest clusters in our simulations.  
As described in Section~\ref{sec:methods}, by default we solve the quadratic programming problem with the normalization constraint (equation~\eqref{eq:inte_f_constraint}) and the non-negativity ($f\ge0$) constraint, but will discuss results without these constraints.

To compute the model power spectrum given a non-parametric $f$, we assign each particle in a simulation with an $f$ weight according to its associated $(\delta_1, \nabla^2\delta_1, \mathcal{G}_2)$ values in the initial space.  We then grid the particles onto a $512^3$ grid using CIC interpolation, which gives a Nyquist frequency of $0.8\ h\ {\rm Mpc}^{-1}$.  We choose to use this finer grid when computing the halo power spectrum to avoid aliasing effects \cite{jing05}.

\subsection{Results using large boxes}
\label{sec:results_f_and_P}

Here we derive the $f$ solutions for 6 simulations and compute their mean.  We then apply the average $f$ to 10 different simulations to calculate the model power spectra, their mean, and the error-bars on the power.
In this way the first 6 simulations act as the training set, whereas the latter 10 serve as a cross-validation set.
This shows that our $f$ is not overfitting to specific details of the simulations.

\subsubsection{Physical quantities that best describe the halo field}
\label{sec:results_N150}

\begin{figure}[tbp]
\centering
\includegraphics[width=0.7\linewidth]{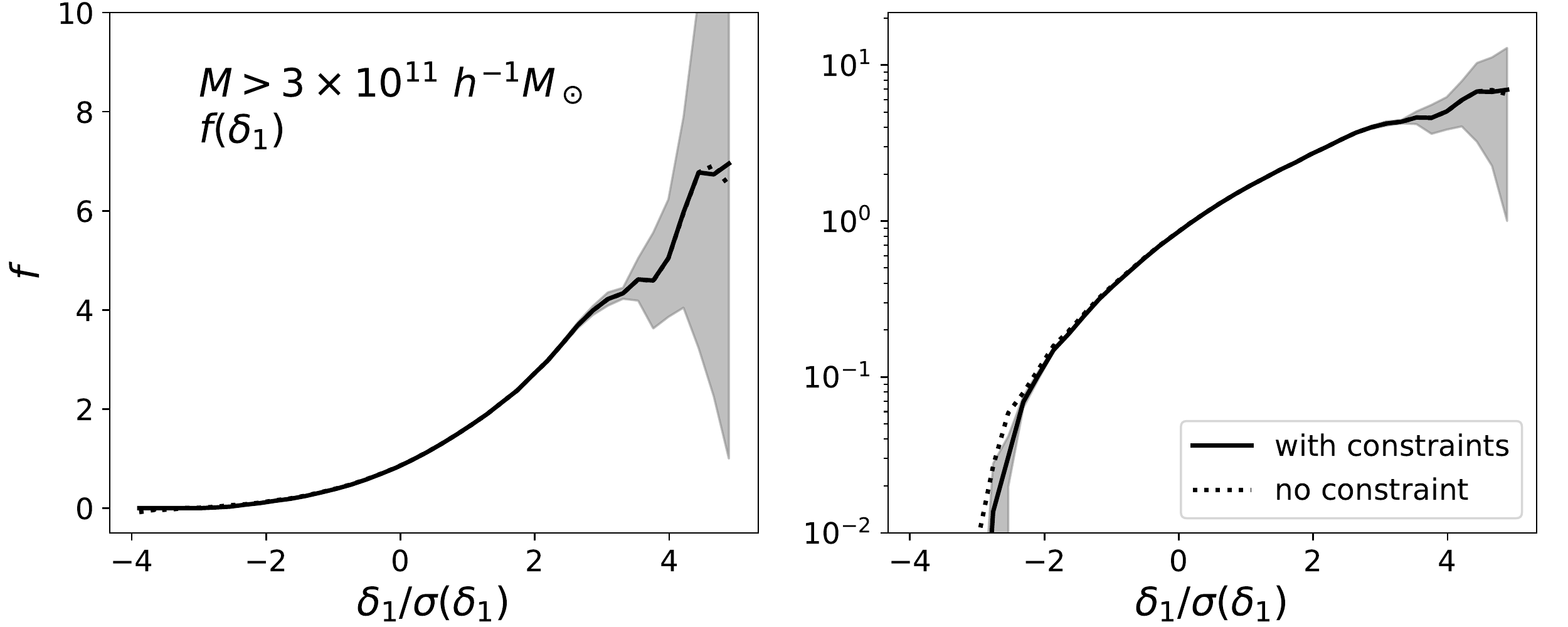}
\includegraphics[width=\linewidth]{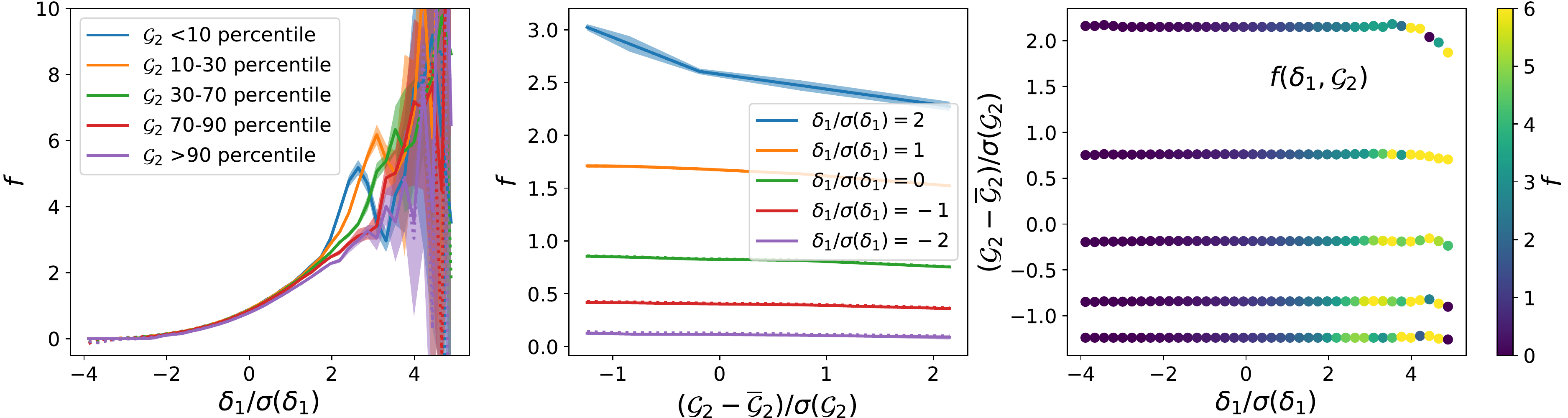}
\includegraphics[width=\linewidth]{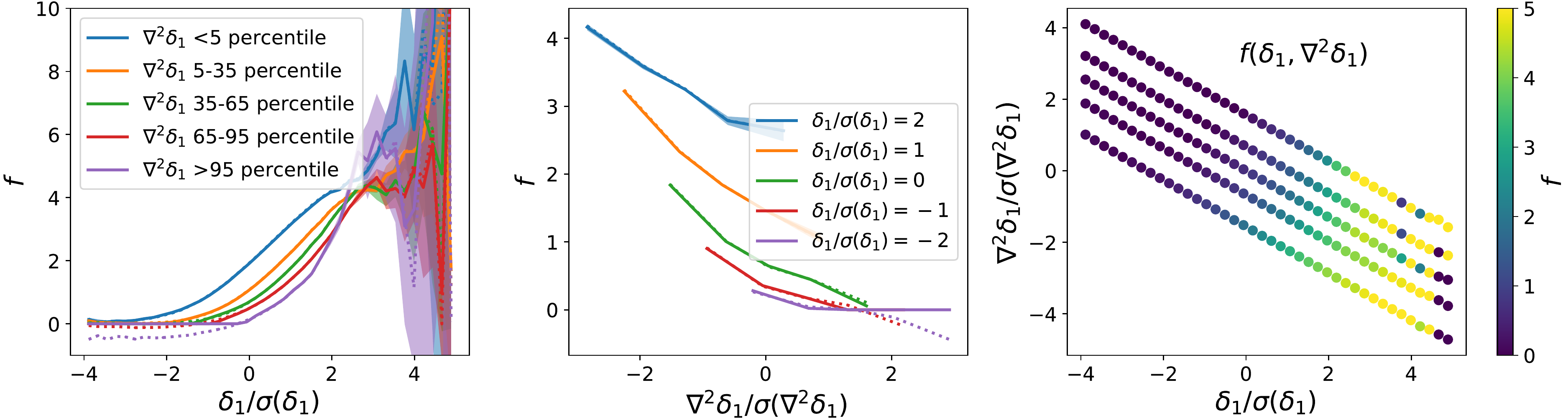}
\caption{\label{fig:f_N150} The halo-to-mass ratios ($f$) fitted for mass-weighted halos with $M>3\times10^{11}\ h^{-1}\ M_\odot$, assuming that $f$ depends on different physical quantities.
From top to bottom: $f$ (defined in equation~\eqref{eq:delta_model_0}) as a function of $\delta_1$, $(\delta_1, \mathcal{G}_2)$, and $(\delta_1, \nabla^2\delta_1)$.  The function $f$ is obtained using $R_f=3\ h^{-1}$~Mpc and $k_{\rm max}=0.3\ h\ {\rm Mpc}^{-1}$ on $400^3$ grids, and is averaged over 6 simulations.  Shades represent 1-$\sigma$ scatter.  Solid lines show solutions of $f$ with the normalization constraint (equation~\eqref{eq:inte_f_constraint}) and the $f\ge0$ constraint, while dotted lines represent solutions without constraints.  The left and right panels of the top row show the same $f$ in linear and log scales respectively.  The right panel of the middle row illustrates $f$ in the 2-dimensional $(\delta_1, \mathcal{G}_2)$ plane.  The left panel shows $f$ as a function of $\delta_1$ in different (percentile) bins of $\mathcal{G}_2$, and the middle panel presents $f$ as a function of $\mathcal{G}_2$ at different values of $\delta_1$.  The bottom row is similar to the middle one, but showing results for $f(\delta_1, \nabla^2\delta_1)$.  We note that the mean of $\nabla^2\delta_1$ is zero, while the mean of $\mathcal{G}_2$ is not.  Effects of including constraints are more evident for $f(\delta_1, \nabla^2\delta_1)$.}
\end{figure}

We first focus on modeling the mass-weighted halo field with a mass threshold of $M>3\times10^{11}\ h^{-1}\ M_\odot$ at $z=0.5$.  Figure~\ref{fig:f_N150} shows $f$ averaged over the 6 training simulations.  The top, middle, and bottom rows illustrate $f$ as a function of $\delta_1$, $(\delta_1, \mathcal{G}_2)$, and $(\delta_1, \nabla^2\delta_1)$ respectively.  Shades represent 1-$\sigma$ scatter.  Solid and dotted lines show solutions of $f$ with and without constraints.  The left and right panels of the top row show $f$ in linear and log scales respectively.  The right panel of the middle row illustrates $f$ in the 2-dimensional $(\delta_1, \mathcal{G}_2)$ plane.  The left panel shows $f$ as a function of $\delta_1$ in different (percentile) bins of $\mathcal{G}_2$, and the middle panel presents $f$ as a function of $\mathcal{G}_2$ at different values of $\delta_1$.  The bottom row is similar to the middle one, but showing results for $f(\delta_1, \nabla^2\delta_1)$.

In all 3 fitting choices with the mass-weighted halos, our non-parametric $f$ obtained with constraints is non-negative and monotonically increasing with $\delta_1$, except in $\delta_1\gtrsim3$ regions where the solution becomes noisy.  The shape of $f$ deviates from a linear or quadratic function of $\delta_1$ as seen from the top right panel of Figure~\ref{fig:f_N150}.  These trends are even more evident with higher halo mass cuts as we will show below.  Such behavior contradicts the prediction from bias expansion, as we will demonstrate later.

While $f$ seems only weakly dependent on $\mathcal{G}_2$, being slightly larger at smaller $\mathcal{G}_2$ when $\delta_1 \gtrsim 2$, it strongly depends on $\nabla^2\delta_1$, showing a clear separation of $f$ in different $\nabla^2\delta_1$ bins.  This implies a small contribution of $\mathcal{G}_2$ in recovering the halo field for the $M>3\times10^{11}\ h^{-1}\ M_\odot$ halos, but a more significant role of $\nabla^2\delta_1$.
We have verified that this conclusion and the trend of variation in $f(\delta_1, \mathcal{G}_2)$ hold for all mass bins considered in this work ($M>3\times10^{11} - 6\times10^{12}\ h^{-1}\ M_\odot$), which we will come back to in Section~\ref{sec:results_diffN}.
The function $f(\delta_1,\nabla^2\delta_1)$ appears to be sensitive to the normalization and non-negativity constraints.  Without constraints, $f$ becomes negative when $\delta_1<0$ which is unphysical, but this trend is mild for $R_f=3\ h^{-1}$~Mpc and $k_{\rm max}=0.3\ h\ {\rm Mpc}^{-1}$.  We will summarize the effects of the non-negativity and normalization constraints in Section~\ref{sec:results_qp_constraints}.  We also note a tendency of $f$ to become non-monotonic with $\nabla^2\delta_1$ at $\delta_1 \gtrsim 2$, both with and without constraints.  These trends become more evident with smaller $R_f$ and larger $k_{\rm max}$, which we will discuss in Section~\ref{sec:results_Rf_and_kmax}.

\begin{figure}[tbp]
\centering
\includegraphics[width=\linewidth]{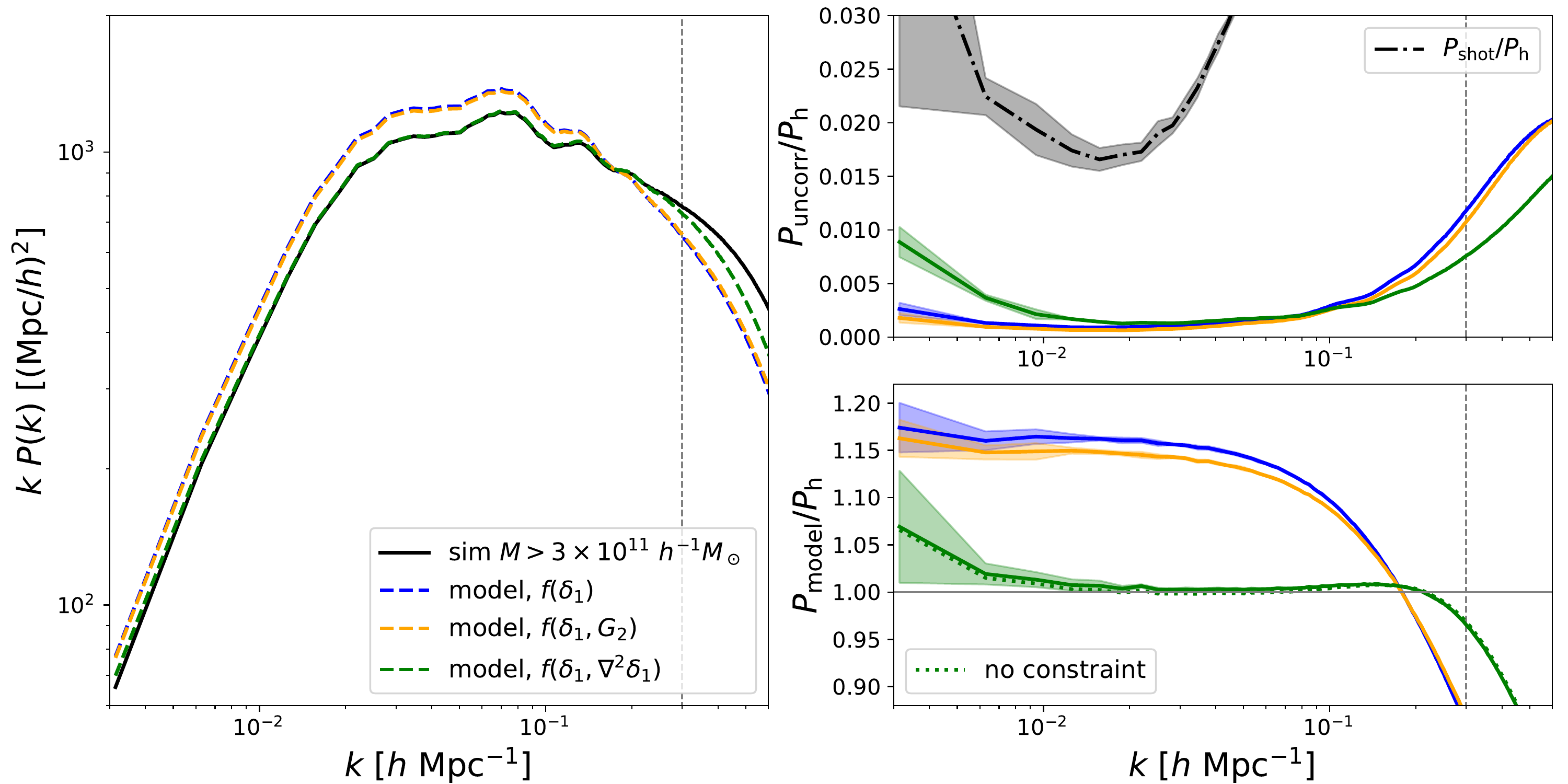}
\caption{\label{fig:P_N150} Recovery of the halo power spectrum of mass-weighted halos with $M>3\times10^{11}\ h^{-1}\ M_\odot$, using the $f$ solutions shown in Figure~\ref{fig:f_N150}.  The power spectra and their ratios are calculated using $512^3$ grids and averaged over 10 simulations that are different from those used for computing $f$.  In each panel the vertical gray dashed line illustrates the $k_{\rm max}$ that we use for fitting.
The black solid line in the left panel shows the measured halo power spectrum $P_{\rm h}$ from the simulations.  Blue, orange, and green dashed lines represent the model power spectra $P_{\rm model}$ using the fitted $f(\delta_1)$, $f(\delta_1,\mathcal{G}_2)$, and $f(\delta_1,\nabla^2\delta_1)$ respectively.
The top right panel illustrates the ratio of the power spectrum of the uncorrelated residual $P_{\rm uncorr}$ to $P_{\rm h}$.  The black dot-dashed line represents the ratio of the conventional Poissonian shot noise to $P_{\rm h}$, shown for illustration purposes only.  Shades show 1-$\sigma$ scatter between simulations of the ratios.
The bottom right panel illustrates $P_{\rm model}/P_{\rm h}$, and the green dotted line represents $P_{\rm model}$ using $f(\delta_1,\nabla^2\delta_1)$ without the normalization and non-negative constraints.}
\end{figure}

Using the $f$ solutions shown above, we assess how well our model recovers the halo power spectrum $P_h$ by calculating the model grid $\delta^{\rm model}_h$ and the model power spectrum $P_{\rm model}$.  We divide $\delta^{\rm model}_h$ into a part that is correlated with $\delta^{\rm true}_h$, and another uncorrelated residual part.  The power spectrum of the uncorrelated residual is \cite{modi17}
\begin{equation}
P_{\rm uncorr} = P_{\rm model} - P_{\rm h,model}^2/P_{\rm h},
\end{equation}
where $P_{\rm h,model}$ is the cross power spectrum of the halo and model grids.  $P_{\rm uncorr}$ thus acts as a metric of the quality of the fit, and $P_{\rm model} - P_{\rm uncorr} = P_{\rm h,model}^2/P_{\rm h}$ gives the power spectrum of the correlated part.

Figure~\ref{fig:P_N150} compares the model power spectra $P_{\rm model}$ computed using the $f$ solutions shown above to the measured halo power spectrum $P_{\rm h}$ from the simulations.
The vertical gray dashed lines mark $k_{\rm max}$.
The black solid line in the left panel shows $P_{\rm h}$.  Blue, orange, and green dashed lines represent $P_{\rm model}$ using the fitted $f(\delta_1)$, $f(\delta_1,\mathcal{G}_2)$, and $f(\delta_1,\nabla^2\delta_1)$ respectively.
The top right panel illustrates the ratio $P_{\rm uncorr}/P_{\rm h}$.  Shades show 1-$\sigma$ scatter between simulations of the ratios.
The bottom right panel illustrates a second metric of the goodness of fit of our model, $P_{\rm model}/P_{\rm h}$, and the green dotted line represents $P_{\rm model}$ using $f(\delta_1,\nabla^2\delta_1)$ without the normalization and non-negative constraints.

The black dot-dashed line in the top right panel of Figure~\ref{fig:P_N150} represents the ratio of the conventional shot noise power spectrum $P_{\rm shot}$ to $P_{\rm h}$, where $P_{\rm shot}$ is calculated assuming Poisson sampling and mass weighting \cite{seljak09}.  We note that we did not subtract this conventional shot noise in our analysis, and $P_{\rm shot}$ is only shown for illustration purposes.  Any irreducible randomness in $\delta^{\rm model}_h$ should appear in the uncorrelated part with $\delta^{\rm true}_h$ by construction, and therefore be reflected in $P_{\rm uncorr}$.  However, $P_{\rm uncorr}$ is about a factor of 10 smaller than $P_{\rm shot}$, indicating that the uncorrelated residual in $\delta^{\rm model}_h$ is well below the Poisson expectation.  Our results also agree with \cite{seljak09} in that mass weighting suppresses shot noise.

Clearly, the model with $f(\delta_1,\nabla^2\delta_1)$ results in the best recovery of the halo power spectrum, with $P_{\rm model}$ matching $P_{\rm h}$ at the $0.5\%$ level from $0.01\ h\ {\rm Mpc}^{-1}$ to $k_{\rm max}=0.3\ h\ {\rm Mpc}^{-1}$.  
The other two models, $f(\delta_1)$ and $f(\delta_1,\mathcal{G}_2)$, on the other hand, lead to over 15\% overestimates of $P_{\rm model}/P_{\rm h}$.  This overestimation remains at 13\% when considering both $\delta_1$ and $\mathcal{G}_2$.
We thus find that $\nabla^2\delta_1$ is key to modeling the mass-weighted halo field.  We note that our model with $\delta_1$ and $\nabla^2\delta_1$ overestimates $P_{\rm h}$ at $k<0.01\ h\ {\rm Mpc}^{-1}$ by up to 5\%, but the power spectrum of the uncorrelated residual $P_{\rm uncorr}$ is less than 1\% of $P_{\rm h}$ at these wavenumbers.  This indicates that the overestimation is driven by the component of $\delta_h^{\rm model}$ that is correlated with $\delta^{\rm true}_h$, rather than the uncorrelated residuals.  Moreover, the $f(\delta_1,\nabla^2\delta_1)$ solution with constraints raises the amplitude of $P_{\rm model}$ by $0.5\%$ compared to the solution without constraints, which we will come back to in Section~\ref{sec:results_qp_constraints}.

Noticeably, although $f(\delta_1,\nabla^2\delta_1)$ leads to the best recovery of the halo power spectrum, $f(\delta_1,\mathcal{G}_2)$ results in the lowest amplitude of $P_{\rm uncorr}$ instead.  The reason why $f(\delta_1,\nabla^2\delta_1)$ leads to the highest $P_{\rm uncorr}$ is unclear to us.  We note that we do not have an errorbar for the halo field in each grid cell, so we do not have a $\chi^2$ value for the goodness of fit.

For the rest of the paper we will only focus on results using $f(\delta_1, \nabla^2\delta_1)$ with the normalization and non-negativity constraints.

\subsubsection{Results with different halo mass cuts}
\label{sec:results_diffN}

\begin{figure}[tbp]
\centering
\includegraphics[width=\linewidth]{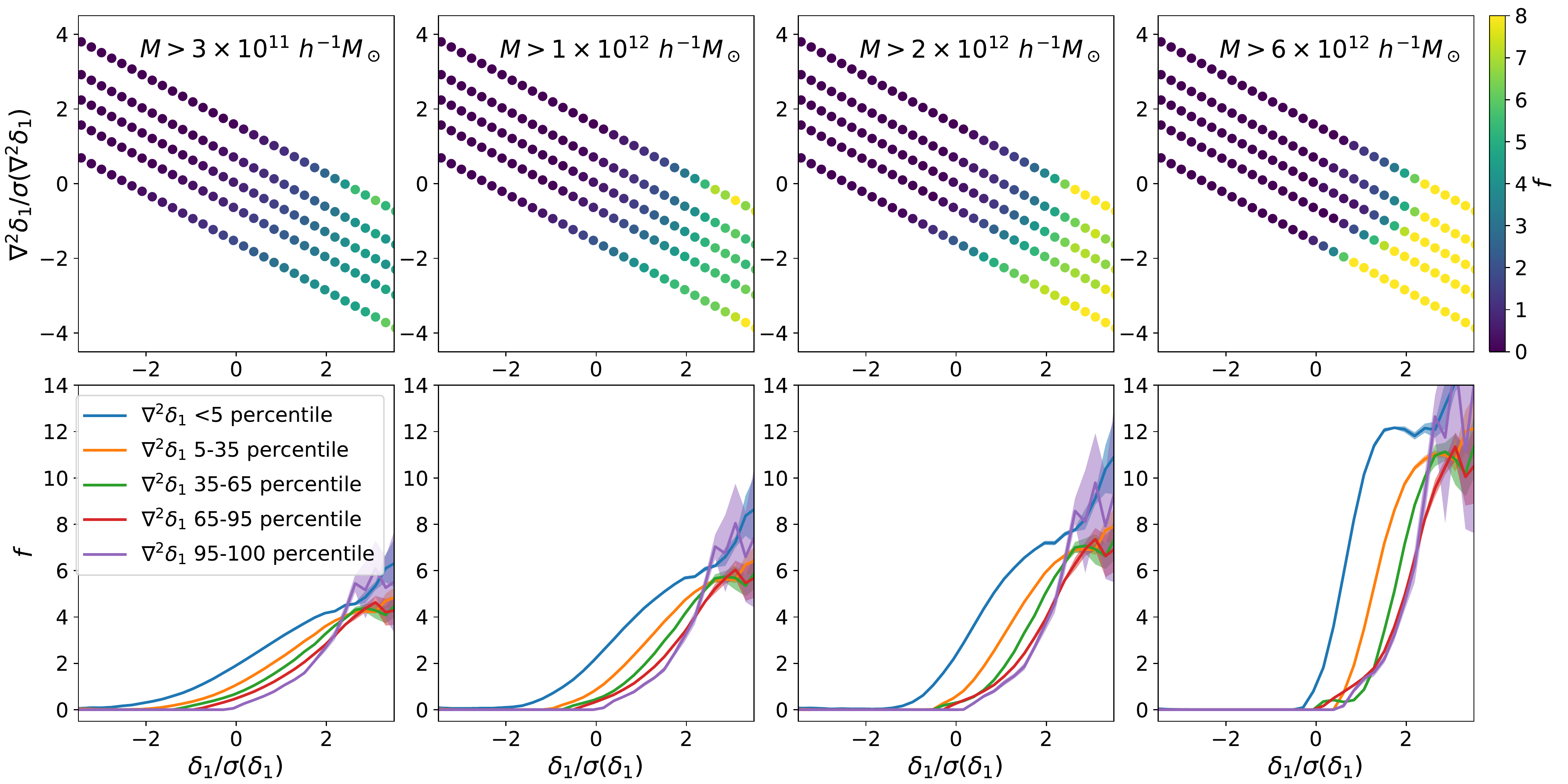}
\caption{\label{fig:f_diffN} The fitted $f(\delta_1, \nabla^2\delta_1)$ for different mass thresholds, obtained using $R_f=3\ h^{-1}$~Mpc and $k_{\rm max}=0.3\ h\ {\rm Mpc}^{-1}$ on $400^3$ grids, and averaged over 6 simulations.  From left to right: $M>3\times10^{11}, 1\times10^{12}, 2\times10^{12}, 6\times10^{12}\ h^{-1}\ M_\odot$.  Top panels show $f$ in the 2-dimensional $(\delta_1, \nabla^2\delta_1)$ plane, and bottom panels illustrate $f$ as a function of $\delta_1$ in different $\nabla^2\delta_1$ bins.  Shades represent 1-$\sigma$ scatter.}
\end{figure}

We now discuss the results of fitting the halo field with 4 different mass cuts $M>3\times10^{11}, 1\times10^{12}, 2\times10^{12}, 6\times10^{12}\ h^{-1}\ M_\odot$, which will test the robustness of our method for a broad range of halo masses.  Figure~\ref{fig:f_diffN} illustrates the averaged $f(\delta_1,\nabla^2\delta_1)$ for these halo mass cuts.  Top panels show $f$ in the 2-dimensional $(\delta_1, \nabla^2\delta_1)$ plane, and bottom panels illustrate $f$ as a function of $\delta_1$ in different $\nabla^2\delta_1$ bins.
Larger halo mass cuts lead to more evident deviation of $f$ from a polynomial of $\delta_1$ and $\nabla^2\delta_1$.  Especially for $M>6\times10^{12}\ h^{-1}\ M_\odot$ halos, $f$ soars up at $\delta_1>0$ and shows large gradients with $\nabla^2\delta_1$.  
This reflects that higher mass halos are exponentially rarer and their formation depends more heavily on the density peaks.
The solutions also gradually become non-monotonic in $\nabla^2\delta_1$ for higher halo mass cuts, suggesting a potential failure of the model and a need for larger $R_f$ to fit the more massive halos.  We have verified that $f(\delta_1,\mathcal{G}_2)$ does not lead to better recovery of the halo power spectrum even for our largest halo mass threshold $6\times10^{12}\ h^{-1}\ M_\odot$.
We note that most previous works on the tidal shear bias studied more massive halos than ours ($M\gtrsim10^{13}\ h^{-1}\ M_\odot$, \cite{baldauf12, chan12, saito14, modi17, abidi18, lazeyras18}, except \cite{bel15, castorina16}). Our findings are broadly consistent with past works, that either the tidal bias is only important for halos with $M\gtrsim10^{13}\ h^{-1}\ M_\odot$ \cite{abidi18}, or there is a small negative shear bias regardless of halo mass \cite{bel15, lazeyras18}, since our $f$ decreases with increasing $\mathcal{G}_2$ (see, however, \cite{modi17} for a different prediction).

\begin{figure}[tbp]
\centering
\includegraphics[width=\linewidth]{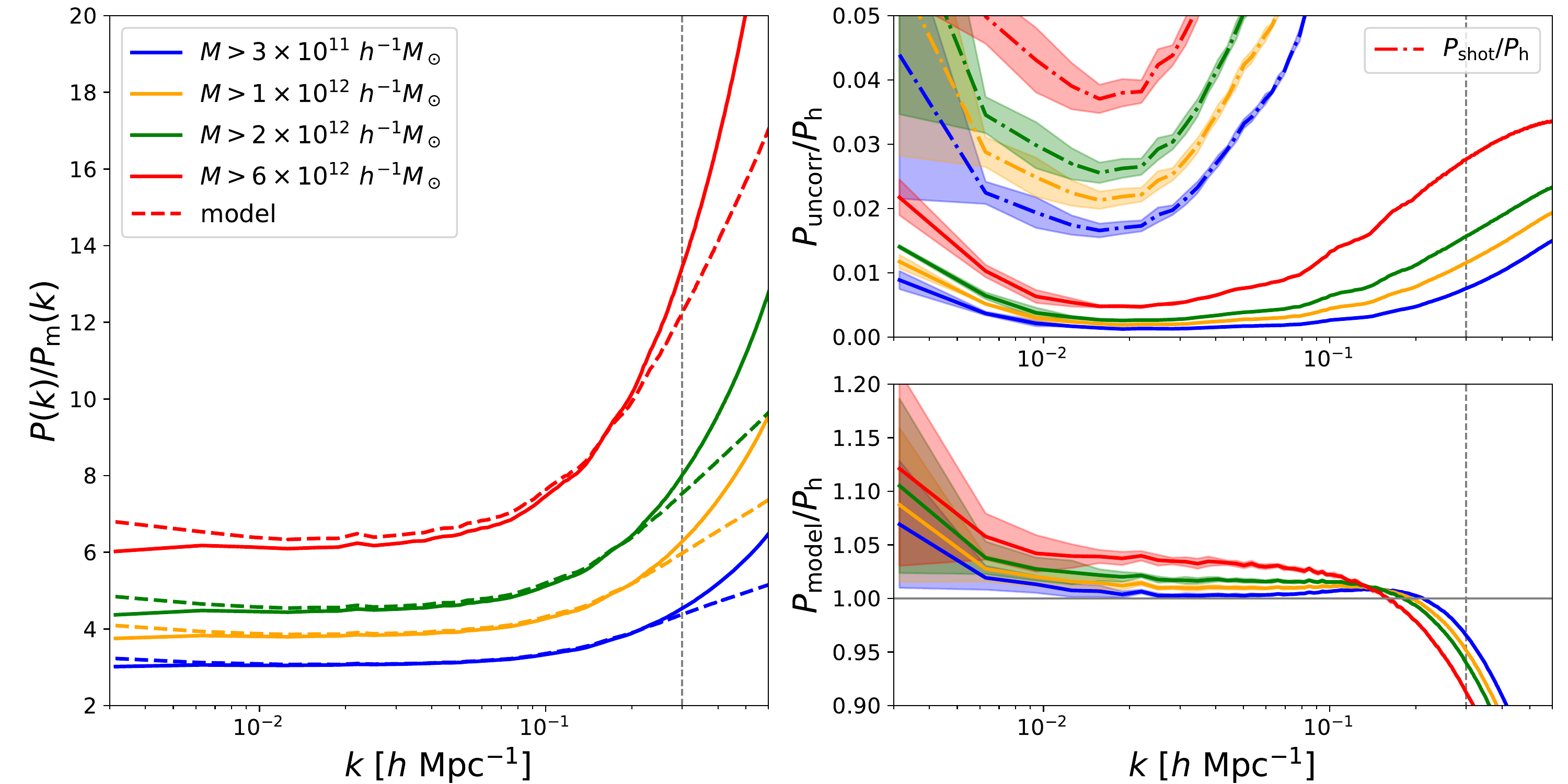}
\caption{\label{fig:P_diffN} Recovery of the halo power spectrum of mass-weighted halos with $M>3\times10^{11}, 1\times10^{12}, 2\times10^{12}, 6\times10^{12}\ h^{-1}\ M_\odot$, represented by blue, orange, green, and red colors respectively.  In each panel the vertical gray dashed line illustrates the $k_{\rm max}$ that we use for fitting.  Left panel presents the ratio of halo power spectra to the matter power spectrum.  Solid and dashed lines represent $P_{\rm h}$ and $P_{\rm model}$ respectively, where $P_{\rm model}$ is obtained using the $f(\delta_1,\nabla^2\delta_1)$ solutions shown in Figure~\ref{fig:f_diffN}.  Dot-dashed lines in the top right panel illustrate $P_{\rm uncorr}/P_{\rm h}$, and the dot-dashed lines represent the ratio of the conventional Poissonian shot noise to $P_{\rm h}$, for illustration purposes only}.  Bottom right panel shows $P_{\rm model}/P_{\rm h}$.  Shades represent 1-$\sigma$ scatter between simulations.
\end{figure}

Figure~\ref{fig:P_diffN} shows the halo power spectra with the 4 mass cuts using the $f$ solutions presented above.  The left panel shows the halo power spectra divided by the matter power spectrum, which at low $k$ gives the linear bias.
The departure from a constant bias is visible for $k\gtrsim 0.1\ h\ {\rm Mpc}^{-1}$.
The two right panels illustrate $P_{\rm uncorr}/P_h$ and $P_{\rm model}/P_h$.
As we mentioned in Section~\ref{sec:results_N150}, $P_{\rm uncorr}$ contains all the information of the uncorrelated residual in $\delta^{\rm model}_h$ with $\delta^{\rm true}_h$.  The conventional Poissonian calculation of the shot noise does not apply in our situation and is only shown for illustration purposes.

For mass thresholds up to $2\times10^{12}\ h^{-1}\ M_\odot$, our $f(\delta_1, \nabla^2\delta_1)$ solutions reproduce the halo power spectra to within 2\% error from $k=0.01\ h\ {\rm Mpc}^{-1}$ to $k_{\rm max}$.  However, halos with higher mass cuts experience an overestimation of $P_{\rm model}/P_{\rm h}$, with the $M>6\times10^{12}\ h^{-1}\ M_\odot$ case seeing a 4\% shift.  
We will discuss the dependence of the modeling results on $R_f$ and $k_{\rm max}$ in Section~\ref{sec:results_Rf_and_kmax} and how a slight increase in $R_f$ might mitigate the overestimation problem with the largest halo mass cut.

For all halo mass cuts, at the low-$k$ end $P_{\rm model}$ overestimates $P_{\rm h}$ by about 5\% more than in the intermediate $k$ range of $0.01-0.1\ h\ {\rm Mpc}^{-1}$.
This issue is unlikely to be caused by the uncorrelated residual part in $\delta_h^{\rm model}$, as $P_{\rm uncorr}$ is less than 1-2\% of $P_{\rm h}$ at low $k$, unable to explain the 5\% overestimation.  We conjecture that the issue is partly caused by $\nabla^2\delta_1$ lacking support at low $k$, given that its Fourier transform goes as $-k^2$ times that of $\delta_1$.  However, $\nabla^2\delta_1$ also carries information of the smoothing scale through the gradient operator.  We thus speculate that incorporating multiple smoothing scales in the modeling, especially one with large $R_f$, might mitigate this low-$k$ overestimation of the halo power spectrum.  We leave an exploration of this question for future work, as the drastically increasing number of dimensions in $f$ when including more smoothing scales makes it hard to solve the problem with quadratic programming.  Machine learning may provide a better approach to this.

Finally, we point out that since we use different sets of simulations to compute $f$ and apply to $P_{\rm model}$, this shows that the $f$ solutions can be cross-validated well across simulations.  We will further demonstrate that applying the $f$ solutions from $500\ h^{-1}$~Mpc small box simulations to the $2\ h^{-1}$~Gpc large boxes also results in good matches of $P_{\rm model}$ to $P_{\rm h}$.

\subsubsection{Comparison with EPS}
\label{sec:results_EPS}

We now show that the $f(\delta_1)$ functions we obtained from the simulation agree qualitatively with analytic predictions from the extended Press-Schechter (EPS) formalism.  
Supposing that $f$ depends on $\delta_1$ only, EPS states that the function $f$ that modulates the amount of mass collapsed into halos more massive than a given threshold $M$ is given by \cite{matsubara08}
\begin{equation}
f(\delta_1) = \frac{\int_M^\infty M n(M', z | \delta_1, R_f) \mathrm{d}M'}{\int_M^\infty M n(M', z) \mathrm{d}M'},
\end{equation}
where $n(M,z)$ is the halo mass function, and $n(M,z | \delta_1,R_f)$ represents the conditional mass function, which gives the number density of halos of mass $M$, identified at redshift $z$, in a region of Lagrangian radius $R_f$ in which the linear overdensity extrapolated to the present time is $\delta_1$.  In our case, instead of a real-space top-hat filter with radius $R_f$, we adopt a Gaussian filter.  For illustrative purposes, we calculate $f$ using the Press-Schechter mass function, which gives
\begin{equation}
f(\delta_1) = \frac{{\rm erfc}\left( \left( \delta_c(z) - \delta_1 \right) / \left( 2 \sqrt{\sigma^2(M) - \sigma^2_{R_f}} \right) \right)}{{\rm erfc}\left( \delta_c(z) / 2 \sigma(M) \right)},
\label{eq:eps_f}
\end{equation}
where $\delta_c(z)$ is the critical overdensity required for spherical collapse at redshift $z$, $\sigma(M)$ is the variance of mass overdensity in a spherical region of size corresponding to the mass scale $M$ and is linearly extrapolated to $z=0$, and $\sigma_{R_f}$ is the mass variance in a Gaussian filter with size $R_f$.  For reference, evaluating the first and second derivatives of $f$ at $\delta_1=0$ gives the bias expansion $f(\delta_1) = f(0) + b_1\delta_1 + b_2\delta_2^2 / 2!$ (though in Section~\ref{sec:results_bias} we will fit for $b_1, b_2$ instead of taking a numerical derivative of $f$).

\begin{figure}[tbp]
\centering
\includegraphics[width=\linewidth]{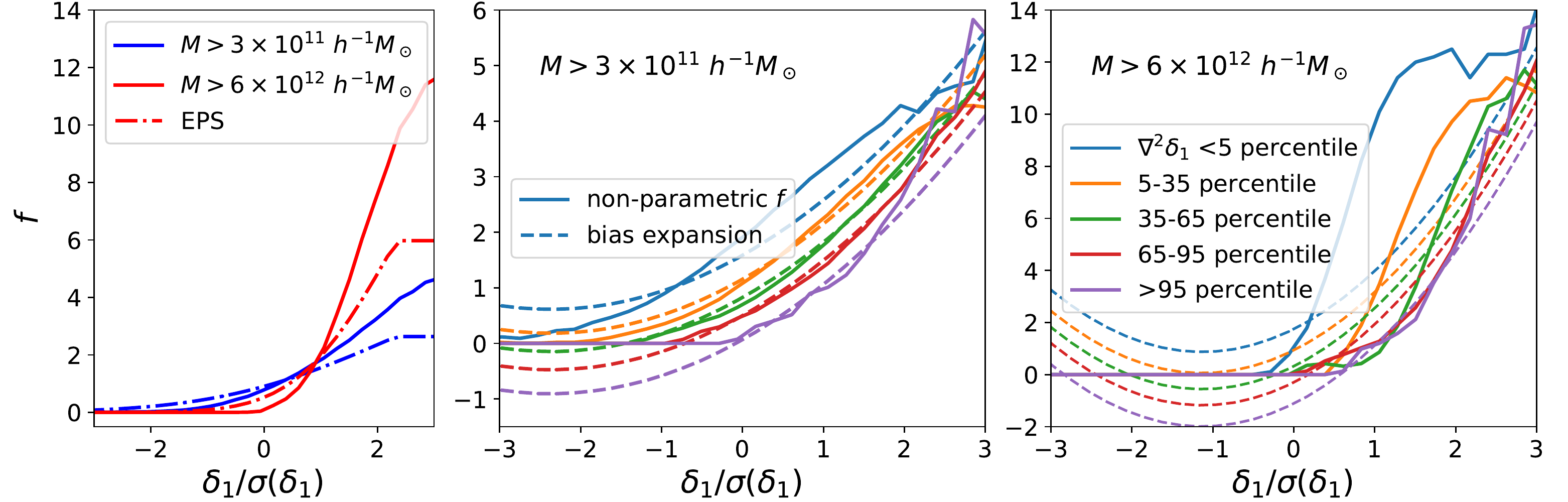}
\caption{\label{fig:f_diffN_bias} Comparison of our non-parametric $f$ to the predictions from EPS and the bias expansion for two halo mass cuts.
Left panel: comparison of the EPS $f$ (equation~\eqref{eq:eps_f}, dot-dashed lines) and the non-parametric $f$ (solid lines).  The latter takes the weighted averaged of $f(\delta_1, \nabla^2\delta_1)$, with the weights given by the percentile ranges of the $\nabla^2\delta_1$ bins.  Blue and red represent halo mass cuts of $M>3\times10^{11}$ and $6\times10^{12}\ h^{-1}\ M_\odot$ respectively.  The middle and right panels compare the non-parametric $f$ (solid lines) to the best-fit bias expansion (dashed lines) in different $\nabla^2\delta_1$ bins represented by different colors (following Figure~\ref{fig:f_diffN}), for the $M>3\times10^{11}$ and $6\times10^{12}\ h^{-1}\ M_\odot$ halos respectively.}
\end{figure}

The left panel of Figure~\ref{fig:f_diffN_bias} compares the EPS prediction of $f(\delta_1)$ (dashed lines) and our non-parametric $f$ (thick solid lines) with two halo mass cuts: $M>3\times10^{11}$ (blue) and $6\times10^{12}\ h^{-1}\ M_\odot$ (red).  The non-parametric $f$ as a function of $\delta_1$ plotted here takes the weighted averaged of $f(\delta_1, \nabla^2\delta_1)$, where the weights are the percentile ranges of the $\nabla^2\delta_1$ bins.  A plateau in the EPS $f$ appears at $\delta_1/\sigma(\delta_1)>2$ because $\delta_1$ is larger than $\delta_c(z)$ at these high $\delta_1$ values, and so the collapsed fraction is saturated at 1, preventing $f$ from growing further.  The overall shape of the EPS $f(\delta_1)$ agrees with our fitted $f$, and our $f$ also shows an indication of flattening at large $\delta_1$ (top right panel of Figure~\ref{fig:f_N150}).

\subsubsection{Comparison with the bias expansion}
\label{sec:results_bias}

We next compare our least-squares non-parametric $f$ to the usual bias expansion.
We assume that $f$ is given by $f(\{\mathcal{O}\}) = \sum_n^{N_{\rm bias}} b_n \mathcal{O}_n$, where $\mathcal{O}_n$ represents the $n$-th operator (out of $N_{\rm bias}$) and $b_n$ is its associated bias parameter.
Starting with equation~\eqref{eq:delta_model_0}, we have
\begin{equation}
1+\delta^{\rm model}_{h, j} = \sum_i w_{ij} f(\{\mathcal{O}\}_i) = \sum_m \sum_{i\in\mathcal{I}_m} w_{ij} \sum_n b_n \mathcal{O}_{n,m},
\end{equation}
where $\left( \mathcal{O}_n \right)_i$ represents $\mathcal{O}_n$ evaluated at the $i$-th particle and $\mathcal{O}_{n,m}$ is the corresponding value if the $i$-th particle falls into the $m$-th bin in the 3-dimensional volume of $\delta_1$-$\nabla^2\delta_1$-$\mathcal{G}_2$.  Rearranging the above equation gives
\begin{equation}
1+\delta^{\rm model}_{h, j} 
= \sum_n \sum_m A_{jm} \underbrace{\mathcal{O}_{n,m}}_{T_{mn}} b_n = \sum_n B_{jn} b_n
\end{equation}
where $T$ is a $N_{\rm bins} \times N_{\rm bias}$ matrix with $T_{mn}$ = $\mathcal{O}_{n,m}$, and $B = AT$.  Therefore the polynomial bias expansion solution $f_{\rm poly} = (b_0, b_1, ..., b_{N_{\rm bias}})$ reads
\begin{equation}
f_{\rm poly} = \left( B^T B \right)^{-1} B^T \left( 1 + \delta^{\rm true}_h \right).
\label{eq:f_bias}
\end{equation}

Following \cite{vlah16, schmittfull19, kokron21}, we expand up to second-order bias
\begin{equation}
f(\{ \mathcal{O} \}) = b_0 + b_1\delta_1 + b_2\delta_1^2 + b_{\nabla^2}\nabla^2\delta_1,
\end{equation}
and use equation~\eqref{eq:f_bias} to calculate the $(b_0, b_1, b_2, b_{\nabla^2})$ coefficients from our simulations, using the $A$ matrices that have already been computed for obtaining the non-parametric $f$.\footnote{Another way of computing the biases is to directly solve the least-squares problem $\sum_n b_n \tilde{\mathcal{O}}_n = 1+\delta_h^{\rm true}$, where $\tilde{\mathcal{O}}_n$ represents the advected fields obtained by CIC interpolating the particles with their corresponding $\mathcal{O}_n$ weights \cite{kokron21}.  We have verified that using our equation~\ref{eq:f_bias} gives roughly the same results as directly fitting the biases.  Since the direct fit method can include all of $\delta_1, \delta_1^2, \nabla^2\delta_1, \mathcal{G}_2$, we have also confirmed that further including $\mathcal{G}_2$ in addition to $\nabla^2\delta_1$ in the bias expansion only affects the recovery of the halo power spectrum by $\lesssim1\%$ for the mass cuts that we considered.  As we mentioned in Section~\ref{sec:results_diffN}, a negligible impact of $\mathcal{G}_2$ for our mass cuts is consistent with previous works on the tidal shear bias.}
The resulting bias expansion solution thus represents fitting the halos at the field level with $(b_0,b_1,b_2,b_{\nabla^2})$, with the same $R_f$ and $k_{\rm max}$ as the non-parametric $f$.  We note that we are still minimizing the real-space squared error, unlike \cite{schmittfull19, kokron21} who essentially minimize the error of fitting the halo power spectrum.  As a consequence, we are not fitting for the power spectrum, but the halo field.
Our approach is also different from \cite{modi17, abidi18} since we do not calculate the bias using the halo-matter cross power spectrum or higher-order halo statistics.  Our definition of the bias and the operators differ from previous works on the bias expansion, but it does not affect the results in this Section and we are more interested in comparing the bias expansion with our non-parametric $f$.

The middle and right panels of Figure~\ref{fig:f_diffN_bias} compare the non-parametric $f$ (solid lines) to the bias expansion $f$ (dashed lines) in different $\nabla^2\delta_1$ bins represented by different colors (as in Figure~\ref{fig:f_diffN}), for the $M>3\times10^{11}$ and $6\times10^{12}\ h^{-1}\ M_\odot$ halos respectively.  For both halo mass cuts, the bias expansion predicts negative $f$ at $\delta_1<0$ in the lower $\nabla^2\delta_1$ bins, which is unphysical.
While the bias expansion roughly captures the shape of the non-parametric $f$ for $M>3\times10^{11}\ h^{-1}\ M_\odot$ (except in the highest $\nabla^2\delta_1$ bin), it completely misses it for $M>6\times10^{12}\ h^{-1}\ M_\odot$.  
This is expected since higher-mass halos showed a steeper behavior with $\delta_1$ in Figure~\ref{fig:f_diffN}, making $f$ less amenable to a bias expansion.
The bias expansion also predicts an unphysically rising $f$ at sufficiently negative overdensities.
\footnote{If we use $R_f=4\ h^{-1}$~Mpc and $k_{\rm max}=0.25\ h\ {\rm Mpc}^{-1}$ for the $M>6\times10^{12}\ h^{-1}\ M_\odot$ halos, which as we will show in Section~\ref{sec:results_Rf_and_kmax} results in the non-parametric $f$ better reproducing $P_{\rm h}$, the resulting non-parametric $f$ becomes slightly more linear but the comparison with the bias expansion remains similar.}

\begin{figure}[tbp]
\centering
\includegraphics[width=\linewidth]{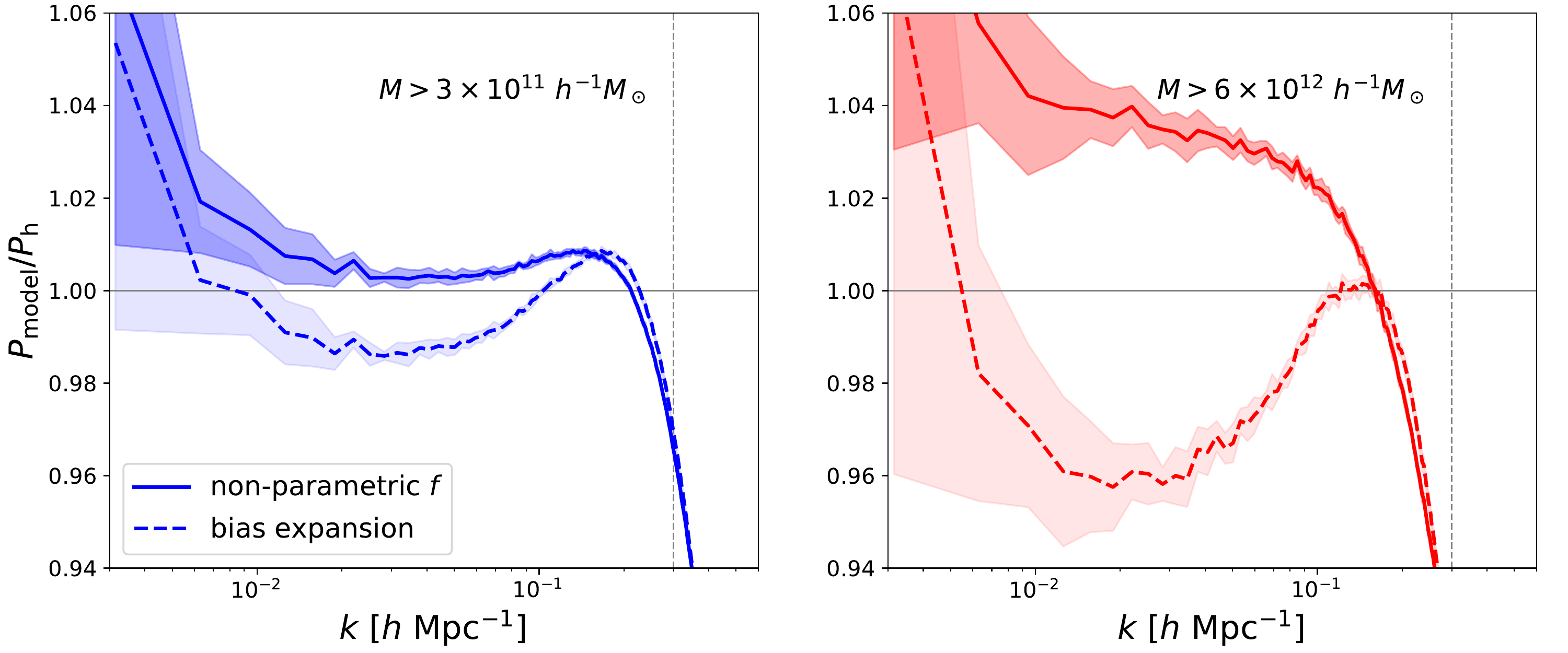}
\caption{\label{fig:P_diffN_bias} Comparison of the power spectra obtained using the non-parametric $f(\delta_1,\nabla^2\delta_1)$ and the bias expansion for mass-weighted halos with $M>3\times10^{11}$ (left) and $6\times10^{12}\ h^{-1}\ M_\odot$ (right).  Solid and dashed lines represent results using the non-parametric $f$ and bias expansion respectively.  Shades represent 1-$\sigma$ scatter.}
\end{figure}

Figure~\ref{fig:P_diffN_bias} compares the power spectra obtained using the non-parametric $f(\delta_1,\nabla^2\delta_1)$ and the bias expansion for mass-weighted halos with $M>3\times10^{11}$ (left) and $6\times10^{12}\ h^{-1}\ M_\odot$ (right).  Solid and dashed lines represent results using the non-parametric $f$ and bias expansion respectively.  The bias expansion underpredicts $P_{\rm model}$ at 1\% and up to 4\% levels at $k=0.01-0.1\ h\ {\rm Mpc}^{-1}$ for $M>3\times10^{11}$ and $6\times10^{12}\ h^{-1}\ M_\odot$ respectively.  The non-parametric $f$ thus outperforms the bias expansion for the $M>3\times10^{11}\ h^{-1}\ M_\odot$ halos since it recovers the halo power spectrum at sub-percent level in the intermediate $k$ range, although it overpredicts the power spectrum of $M>6\times10^{12}\ h^{-1}\ M_\odot$ halos by 3\%.  However, we will show below that using $R_f=4\ h^{-1}$~Mpc and $k_{\rm max}=0.25\ h\ {\rm Mpc}^{-1}$ results in percent level recovery of the power spectrum of the more massive halos, while we have verified that these parameters lead to up to 6\% underestimate of $P_{\rm model}/P_{\rm h}$ when using the bias expansion.

Since we do not fit the power spectrum but the halo field, it may not be surprising that our bias expansion $f$ does not reproduce the halo power spectrum as closely as previous works on Lagrangian biasing \citep[e.g.][]{kokron21}.
Furthermore, we follow a fully Lagrangian approach, not using any $k$-dependent biases or transfer functions on the final (Eulerian) space \citep[e.g.][]{schmittfull19}.
While our 1-4\% underestimation of $P_{\rm model}/P_{\rm h}$ using the bias expansion seems acceptable, we speculate that the span of $f$ in positive and negative values at $\delta_1<0$ likely results in a cancellation of the effects of using such unphysical $f$ that does not preserve monotonicity and non-negativity, and also rises at low $\delta_1$.  An unphysical $f$ that is negative at $\delta_1<0$ may lead to negative halo densities at the final redshift, whose effect may be small on the power spectrum, but is likely important for the one-point function of the halo field.

\subsection{Results using small boxes}
\label{sec:results_smallbox}

Here we derive $f(\delta_1, \nabla^2\delta_1)$ for multiple $500\ h^{-1}$~Mpc small box simulations and apply the solutions to the $2\ h^{-1}$~Gpc large box simulations.  This is motivated by the fact that realistic galaxy populations can only be modeled in small-box (hundreds of Mpc) cosmological hydrodynamical simulations (e.g., Illustris \cite{vogelsberger14}, IllustrisTNG \cite{weinberger17, pillepich18, tng_dr}, EAGLE \cite{schaye15, eagle_dr}, BAHAMAS \cite{mccarthy17}, MAGNETICUM \cite{hirschmann14}, Horizon-AGN \cite{dubois14}), while large box sizes are required to capture the low-wavenumber modes and allow for a systematic exploration of halo clustering.  Moreover, recent studies have begun to explore galaxy bias in cosmological hydrodynamical simulations \cite{chavesmontero16, springle18, monterodorta20, barreira21}.  We thus aim to test whether small-box simulations produce $f$ solutions consistent with the large boxes, and whether these solutions lead to a precise match of $P_{\rm model}$ to $P_{\rm h}$ when applied to large box simulations.

\begin{figure}[tbp]
\centering
\includegraphics[width=\linewidth]{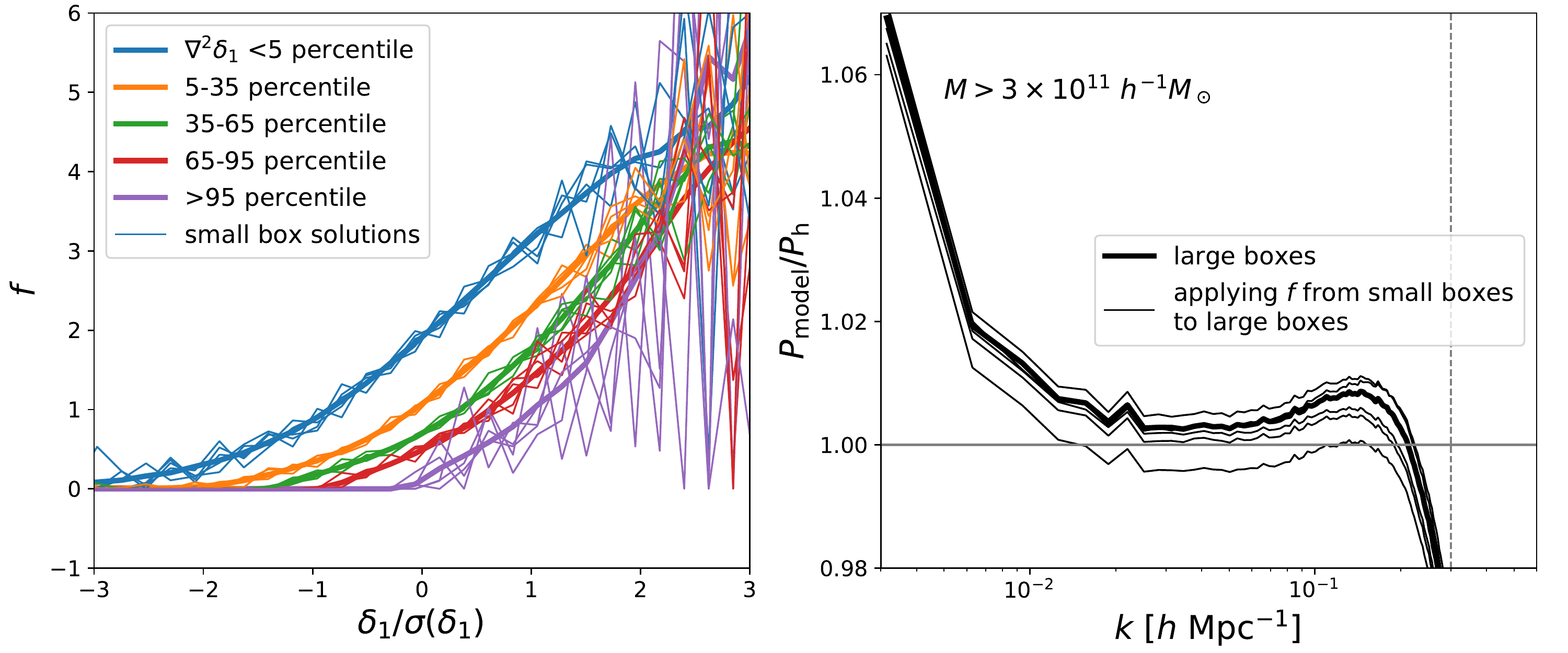}
\caption{\label{fig:f_and_P_N150_small} The $f$ solutions obtained from small-box simulations and the corresponding $P_{\rm model}$, compared to those derived from the large boxes alone.  Left panel compares the $f(\delta_1,\nabla^2\delta_1)$ solutions for the $M>3\times10^{11}\ h^{-1}\ M_\odot$ halos obtained from the $2\ h^{-1}$Gpc boxes (thick lines) to those from five $500\ h^{-1}$~Mpc boxes (thin lines).  Different colors represent different $\nabla^2\delta_1$ bins, as in the bottom left panel of Figure~\ref{fig:f_N150}.  Right panel shows $P_{\rm model}/P_{\rm h}$.  The thick line represents the result of applying $f$ found from the big boxes to other big boxes, whereas the thin lines illustrate the results of using $f$ from the small boxes to the big ones.  Each $P_{\rm model}/P_{\rm h}$ curve is averaged over 10 big-box simulations to reduce Poisson noise.}
\end{figure}

We derive the $f$ solutions for mass-weighted halos within 5 small-box simulations, using the same $R_f=3\ h^{-1}$~Mpc, $k_{\rm max}=0.3\ h\ {\rm Mpc}^{-1}$, and $5\ h^{-1}$~Mpc cell size as the large boxes, therefore $100^3$ grids to interpolate the particles.  Figure~\ref{fig:f_and_P_N150_small} shows the results of our calculations for the $M>3\times10^{11}\ h^{-1}\ M_\odot$ halos.
The thin lines in the left panel illustrate the different $f$ solutions obtained from the small boxes, and the thick lines represent the averaged $f$ solution from the 6 large boxes discussed above.  Different colors represent different $\nabla^2\delta_1$ bins, as in the bottom left panel of Figure~\ref{fig:f_N150}.
The small-box $f$ solutions fluctuate around the large-box ones and have more scatter in the less occupied $\delta_1$ and $\nabla^2\delta_1$ bins.
The right panel shows $P_{\rm model}/P_{\rm h}$, where each curve is averaged over 10 large-box simulations.  The thick line represents using the averaged $f$ from the large boxes, while the 5 thin lines illustrate the results of applying each of the 5 $f$ solution to the large boxes to calculate the model grid.  Although variations exist, applying the small box $f$ to large boxes leads to $P_{\rm model}/P_{\rm h}$ consistent with applying the large box $f$ to within sub-percent level.  In future work we plan to test whether such stability holds when using halo occupation distribution models \cite{yuan18, hadzhiyska20, hadzhiyska21}.

\subsection{Effects of different smoothing scales and wavenumber cuts}
\label{sec:results_Rf_and_kmax}

Here we discuss the effects of using different smoothing radii $R_f$ and cutoff wavenumbers $k_{\rm max}$ on the $f$ values and the model power spectrum.  We only calculate $f$ from one simulation for computational efficiency, and compute the associated model power spectrum using that same simulation.
We perform the calculations using $f(\delta_1,\nabla^2\delta_1)$, but show a weighted averaged $f$ with the weights given by the percentile ranges of the $\nabla^2\delta_1$ bins.
We also present $f$ solutions and the resulting $P_{\rm model}$ without the normalization (equation~\eqref{eq:inte_f_constraint}) and non-negativity ($f\ge0$) constraints, but will discuss the effects of these constraints in detail in Section~\ref{sec:results_qp_constraints}.

\begin{figure}[tbp]
\centering
\includegraphics[width=\linewidth]{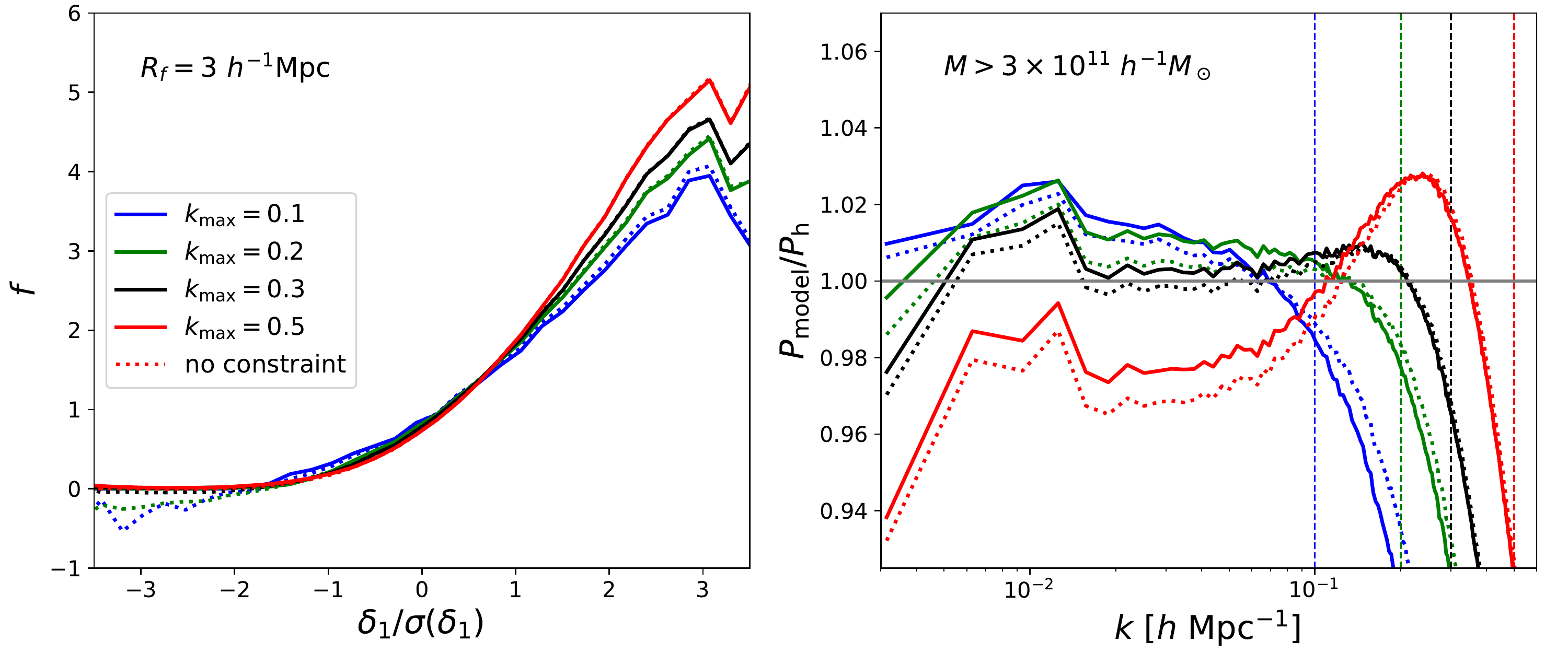}
\caption{\label{fig:f_and_P_diffkmax} Effects of varying $k_{\rm max}$ on $f$ (left) and $P_{\rm model}$ (right), for mass-weighted halos with $M>3\times10^{11}\ h^{-1}\ M_\odot$.  We fix $R_f=3\ h^{-1}$~Mpc.  Blue, green, black, and red represent $k_{\rm max}=0.1,0.2,0.3,0.5\ h\ {\rm Mpc}^{-1}$ respectively.
The left panel illustrates the $\nabla^2\delta_1$-averaged $f$ as a function of $\delta_1$, where solid and dotted lines show results with and without the normalization and non-negativity constraints respectively.  The right panel shows $P_{\rm model}/P_{\rm h}$, and the vertical dashed lines illustrate the corresponding values of $k_{\rm max}$.}
\end{figure}

Figure~\ref{fig:f_and_P_diffkmax} shows the result of varying the $k$ cuts ($k_{\rm max}=0.1,0.2,0.3,0.5\ h\ {\rm Mpc}^{-1}$) on the fit for $f$ (left panel) and $P_{\rm model}/P_{\rm h}$ (right panel) for mass-weighted halos with $M>3\times10^{11}\ h^{-1}\ M_\odot$ using $R_f=3\ h^{-1}$~Mpc.
Solid and dotted lines represent results with and without the normalization and non-negativity constraints respectively, and here we only focus on discussing the former.
We find that $f$ becomes slightly more linear with lower $k$ cuts.  Although not shown in this plot, for the higher $k_{\rm max}=0.5\ h\ {\rm Mpc}^{-1}$, $f$ appears to be more non-monotonic in $\nabla^2\delta_1$ at $\delta_1>2$.
The recovery of the halo power spectrum is best with $k_{\rm max}=0.3\ h\ {\rm Mpc}^{-1}$ since it corresponds to $\sim 1/R_f$.  Lower $k$ cuts raise $P_{\rm model}$ by 1-2\% in the relevant range ($k=0.01\ h\ {\rm Mpc}^{-1} - k_{\rm max}$), while a larger $k_{\rm max}=0.5\ h\ {\rm Mpc}^{-1}$ reduces $P_{\rm model}$ by 2\%.

\begin{figure}[tbp]
\centering
\includegraphics[width=\linewidth]{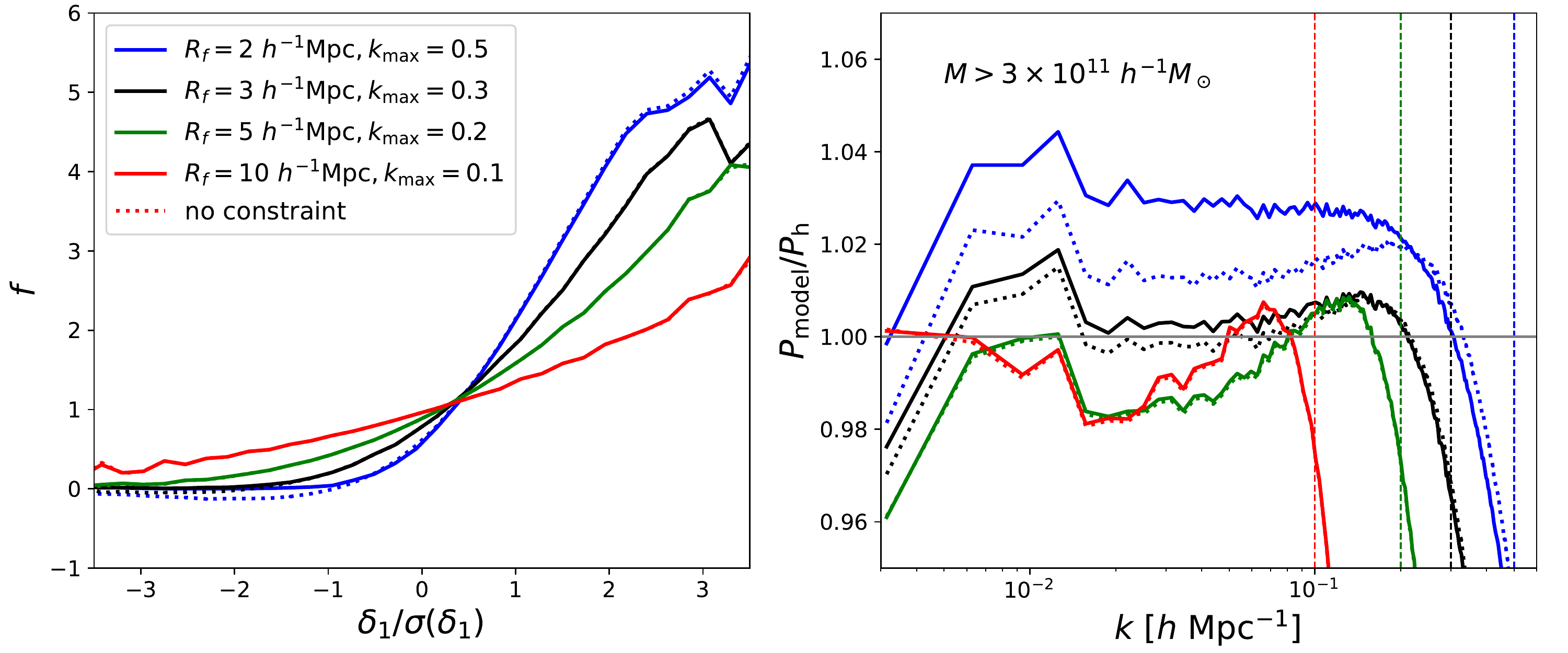}
\caption{\label{fig:f_and_P_diffRf} Effects of different $R_f$ on $f$ and $P_{\rm model}$, for halos with $M>3\times10^{11}\ h^{-1}\ M_\odot$.  Different colors represent varying $R_f$ together with $k_{\rm max}$, similar to Figure~\ref{fig:f_and_P_diffkmax}.}
\end{figure}

Figure~\ref{fig:f_and_P_diffRf} compares the fits for $f$ and $P_{\rm model}/P_{\rm h}$ when using different $R_f$ values for the $M>3\times10^{11}\ h^{-1}\ M_\odot$ halos.  
We choose $R_f=2,3,5,10\ h^{-1}$~Mpc, and set maximum wavenumbers at $k_{\rm max}=0.5,0.3,0.2,0.1\ h\ {\rm Mpc}^{-1}$ for them, where each $k_{\rm max}$ is roughly $1/R_f$. The $f$ solution becomes more linear with larger smoothing scales, consistent with the linear-bias picture.  Setting $R_f=2\ h^{-1}$~Mpc leads to an overestimation of $P_{\rm model}/P_{\rm h}$ by 3\% at $k=0.01\ h\ {\rm Mpc}^{-1} - k_{\rm max}$, while $R_f=5$ and $10\ h^{-1}$~Mpc result in a 1-2\% underestimation.  This points to the need to select a proper smoothing scale according to the mass of the halos under consideration.

\begin{figure}[tbp]
\centering
\includegraphics[width=\linewidth]{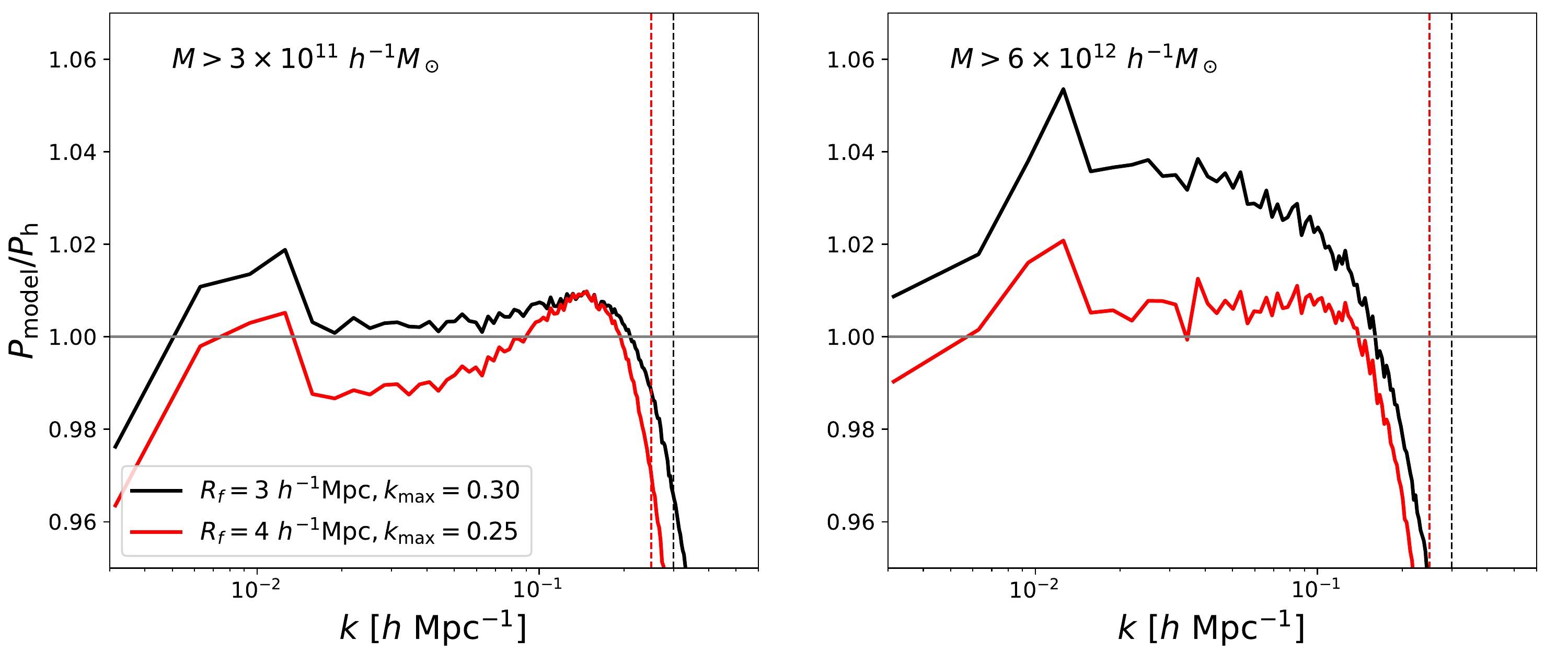}
\caption{\label{fig:diffN_diffRf} 
Ratio $P_{\rm model}/P_{\rm h}$ of the model and halo power spectra for halos with $M>3\times10^{11}\ h^{-1}\ M_\odot$ (left panel) and $6\times10^{12}\ h^{-1}\ M_\odot$ (right panel).  Black and red lines represent results using $R_f=3\ h^{-1}$~Mpc, $k_{\rm max}=0.3\ h\ {\rm Mpc}^{-1}$ and $R_f=4\ h^{-1}$~Mpc, $k_{\rm max}=0.25\ h\ {\rm Mpc}^{-1}$, where the larger $R_f$ provides a better fit for the heavier halos.  The vertical dashed lines illustrate the corresponding values of $k_{\rm max}$.}
\end{figure}

Finally, we explore whether a larger smoothing scale might lead to better recovery of the halo power spectrum for the $M>6\times10^{12}\ h^{-1}\ M_\odot$ halos, as our power-spectrum predictions were biased high in Figure~\ref{fig:P_diffN}.  The mass-weighted average masses of halos above thresholds of $M>3\times10^{11}\ h^{-1}\ M_\odot$ and $6\times10^{12}\ h^{-1}\ M_\odot$ are $2.4\times10^{13}\ h^{-1}\ M_\odot$ and $5.1\times10^{13}\ h^{-1}\ M_\odot$, respectively, which would be the masses contained in Gaussian filters with $R_f=2.6\ h^{-1}$~Mpc and $3.3\ h^{-1}$~Mpc respectively.  Figure~\ref{fig:diffN_diffRf} shows $P_{\rm model}/P_{\rm h}$ for the $M>3\times10^{11}\ h^{-1}\ M_\odot$ (left panel) and $6\times10^{12}\ h^{-1}\ M_\odot$ (right panel) halos.  Black and red lines represent results using $R_f=3\ h^{-1}$~Mpc, $k_{\rm max}=0.3\ h\ {\rm Mpc}^{-1}$ and $R_f=4\ h^{-1}$~Mpc, $k_{\rm max}=0.25\ h\ {\rm Mpc}^{-1}$ respectively.  While $R_f=3\ h^{-1}$~Mpc leads to sub-percent level recovery of the halo power spectrum for the lower mass cut and 3-4\% overestimation of $P_{\rm model}/P_{\rm h}$ for the higher mass cut, $R_f=4\ h^{-1}$~Mpc results in sub-percent recovery of the halo power spectrum for the higher mass cut, but 1\% underestimation of $P_{\rm model}/P_{\rm h}$ for the lower mass cut.  This suggests that it is unlikely to find one smoothing scale that works for all halo mass thresholds, as different mass cuts require a corresponding $R_f$.  We also note that while $R_f=4\ h^{-1}$~Mpc gives a better recovery of the power spectrum of the massive halos, it does not fully eliminate the non-monotonic behavior of $f$ with respect to $\nabla^2\delta_1$, seen in the bottom right panel of Figure~\ref{fig:f_diffN}.

In summary, we find that varying $R_f$ and $k_{\rm max}$ could result in a few percent variations in $P_{\rm model}/P_{\rm h}$ at $k=0.01-0.1\ h\ {\rm Mpc}^{-1}$.  This implies that a sub-percent level of recovery of the halo power spectrum may require fine-tuning of the parameters, although our default choice of $R_f=3\ h^{-1}$~Mpc and $k_{\rm max}=0.3\ h\ {\rm Mpc}^{-1}$ worked out well for $M>3\times10^{11}\ h^{-1}\ M_\odot$ halos.  Future work may explore a broader parameter space and aim to get rid of the fine-tuning.

\subsection{Effects of the non-negativity and normalization constraints}
\label{sec:results_qp_constraints}

We now discuss the effects of including the non-negativity ($f\ge0$) and normalization (equation~\eqref{eq:inte_f_constraint}) constraints on $f$ and $P_{\rm model}$.  The $\nabla^2\delta_1$ bin-averaged $f$ solutions and the corresponding $P_{\rm model}/P_{\rm h}$ without these constraints are shown by the dotted lines in Figure~\ref{fig:f_and_P_diffkmax} and \ref{fig:f_and_P_diffRf}, while Figure~\ref{fig:f_N150} contains a full comparison of $f$ with and without constraints in different $\nabla^2\delta_1$ bins.

Including the constraints affects $f$ primarily at $\delta_1<0$.
When fixing $R_f$ and lowering $k_{\rm max}$ (Figure~\ref{fig:f_and_P_diffkmax}), the $\nabla^2\delta_1$ bin-averaged $f$ shows a tendency of being more negative at $\delta_1/\sigma(\delta_1)<-2$.  While not shown in this plot, for the higher $k_{\rm max}=0.5\ h\ {\rm Mpc}^{-1}$, $f$ is more negative in higher $\nabla^2\delta_1$ bins at $\delta_1/\sigma(\delta_1)<0$, even though the $\nabla^2\delta_1$ bin-averaged $f$ is non-negative.  It also becomes more non-monotonic in $\nabla^2\delta_1$ at $\delta_1/\sigma(\delta_1)>2$.
When changing $R_f$ and setting $k_{\rm max}\sim1/R_f$ (Figure~\ref{fig:f_and_P_diffRf}), the differences in $f$ with and without constraints diminish for larger $R_f$.  For $R_f=2\ h^{-1}$~Mpc, $f$ without constraints becomes more negative in higher $\nabla^2\delta_1$ bins at $\delta_1/\sigma(\delta_1)<0$, which leads to $f<0$ when averaged in $\nabla^2\delta_1$ bins.

The changes in $f$ at $\delta_1<0$ when including the constraints tend to raise $P_{\rm model}$ by up to 2\% at $k=0.01-0.1\ h\ {\rm Mpc}^{-1}$.  This effect seems more evident for the larger $k_{\rm max}=0.5\ h\ {\rm Mpc}^{-1}$ when fixing $R_f=3\ h^{-1}$~Mpc, or when using a small $R_f=2\ h^{-1}$~Mpc with $k_{\rm max}=0.5\ h\ {\rm Mpc}^{-1}$.  For $R_f\ge5\ h^{-1}$~Mpc, $f$ becomes more linear and including the constraints no longer make a big difference in the resulting $f$ and $P_{\rm model}$.  These results echo those of Section~\ref{sec:results_Rf_and_kmax} that our model may still require some fine-tuning of the parameters to well recover the halo power spectrum, which we leave to future work for a detailed exploration.

\section{Conclusions}
\label{sec:conclusions}

We have developed a fully Lagrangian halo biasing model that is non-parametric and qualitatively different from the traditional bias expansion.  We measured the halo-to-mass ratios $f$ using mass-weighted halos in N-body simulations, assuming $f$ is a function of the smoothed linear overdensity $\delta_1$, the tidal operator $\mathcal{G}_2$, and a non-local term $\nabla^2\delta_1$.
Our derived $f$ functions are non-negative and monotonically increasing with $\delta_1$ for mass-weighted halos, unlike a polynomial of $\delta_1$ that does not necessarily guarantee these constraints.  We find that $f$ clearly deviates from a polynomial function of $\delta_1$ as would be expected from the bias expansion.  These trends are more evident for more massive halos, where $f$ starts soaring up at $\delta_1>0$.  
We find that including $\nabla^2\delta_1$ is essential to reproducing the power spectrum of mass-weighted halos.  In particular, our $f(\delta_1,\nabla^2\delta_1)$ is able to recover the power spectrum of mass-weighted halos with $M>3\times10^{11}\ h^{-1}\ M_\odot$ at sub-percent level of accuracy at $k=0.01-0.1\ h\ {\rm Mpc}^{-1}$ given an appropriate smoothing scale to filter the initial density field. 
On the other hand, treating $f$ as a function only of $\delta_1$ leads to a 15\% overestimation of the halo power spectrum.  The inclusion of $\mathcal{G}_2$ only reduces this overestimation by 2\%.  Similar conclusions hold for all halo mass cuts considered in our work, $M>3\times10^{11}-6\times10^{12}\ h^{-1}\ M_\odot$.  This is consistent with previous works which find that either the tidal shear bias is unimportant for halos less massive than $\sim10^{13}\ M_\odot$, or there is a small negative tidal bias across a range of halo masses (\cite{saito14, bel15, abidi18, lazeyras18}, but see also \cite{modi17}).
However, the amplitude of the halo power spectrum is more overestimated with $f(\delta_1,\nabla^2\delta_1)$ for larger halo mass thresholds and at $k<0.01\ h\ {\rm Mpc}^{-1}$, suggesting a need to use larger smoothing scales for more massive halos.
How well the non-parametric $f$ recovers the halo power spectrum is also mildly dependent on input parameters such as the smoothing scale.

By measuring $f(\delta_1,\nabla^2\delta_1)$ using mass-weighted halos in $500\ h^{-1}$~Mpc simulations and applying the resulting $f$ to $2\ h^{-1}$~Gpc simulations, we find that the halo power spectrum can still be matched to within percent level accuracy.  While we have not tested our formalism using number-weighted halos or galaxies populated with a halo-occupation-distribution model, this shows the potential of applying our framework on small-box cosmological hydrodynamical simulations.

We compared our non-parametric $f$ with the $f$ function assuming the bias expansion, which exhibits negative values at $\delta_1<0$ and then rises to positive again at lower overdensities.  We find that using the same smoothing scales and wavenumber cuts, the bias expansion underpredicts the amplitude of the halo power spectrum by up to 4\%.

Having found good performance with our formalism, we list the possible extensions and improvements of our model below:
\setlist{nolistsep}
\begin{itemize}[noitemsep]
\item Test our formalism using number-weighted halos or halos weighted by a halo-occupation distribution;
\item Examine the use of a Poisson likelihood in obtaining $f$ instead of a Gaussian likelihood as in the least-squares fitting, as halo number counts are expected to follow a Poisson distribution;
\item Study whether $\mathcal{G}_2$ plays a more important role in modeling $M>10^{13}-10^{14}\ M_\odot$ halos using our non-parametric model;
\item Apply our model onto high-redshift halos, since halo formation becomes rarer and more extreme at early times;
\item Implement an improved version of our formalism that can take multiple smoothing scales.
\end{itemize}

In summary, we have developed a substantially different picture of describing halo formation compared to the traditional bias expansion approach.  We have also demonstrated a great potential for our non-parametric halo-to-mass ratio to be implemented and tested in future simulations and observational surveys, with some improvements on our formalism in future work.

\acknowledgments

JBM is supported by a Clay fellowship at the Smithsonian Astrophysical Observatory.  DJE is partially supported by U.S. Department of Energy grant DE-SC0013718, NASA ROSES grant 12-EUCLID12-0004, NASA contract NAS5-02015, and as a Simons Foundation Investigator.

\appendix

\paragraph{Note added.} 

\bibliography{References}
\bibliographystyle{JHEP}

\end{document}